\documentclass{emulateapj}
\usepackage{graphics,graphicx,epsfig,lscape}
\usepackage{rotating}
\usepackage{amsmath,amssymb,amsfonts} 
\usepackage{subfigure, multirow}
\usepackage{epstopdf}
\usepackage{natbibspacing}
\usepackage{natbib}
\newcommand{\eg}{\textit{e.g.,}~}
\newcommand{\ie}{\textit{i.e.}~}
\newcommand{\GALEX}{\textit{GALEX}}

\newcommand{\HST}{\textit{HST}}
\newcommand{\Swift}{\textit{Swift}}

\begin{document}

\title{A Search for Lyman Break Galaxies in the CDF-S Using \Swift~UVOT}
\author{
Antara R. Basu-Zych\altaffilmark{1},
Ann E. Hornschemeier\altaffilmark{1},
Erik A. Hoversten\altaffilmark{2},
Bret Lehmer\altaffilmark{3, 4},
Caryl Gronwall\altaffilmark{2}}
\altaffiltext{1}{NASA Goddard Space Flight Center, Code 662, Greenbelt, MD 20771; antara.r.basu-zych@nasa.gov, Ann.Hornschemeier@nasa.gov}
\altaffiltext{2}{Department of Astronomy and Astrophysics, Pennsylvania State University, 525 Davey Laboratory, University Park, PA 16802; hoversten@astro.psu.edu, caryl@astro.psu.edu}
\altaffiltext{3}{Einstein Fellow}
\altaffiltext{4}{Department of Physics and Astronomy, The Johns Hopkins University, 3400 North Charles Street, Baltimore, MD 21218; blehmer@pha.jhu.edu}

\begin{abstract}
While the \Swift~satellite is primarily designed to study gamma-ray bursts, its ultraviolet and optical imaging and spectroscopy capabilities are also being used for a variety of scientific programs. In this study, we use the UV/Optical Telescope (UVOT) instrument aboard \Swift~to discover $0.5<z<2$ Lyman break galaxies (LBGs). UVOT has covered $\sim266$ arcmin$^2$ at $>$60ks exposure time, achieving a limiting magnitude of $u<24.5$, in the Chandra Deep Field South (CDF-S). Applying UVOT near-ultraviolet color selection, we select 50 UV-dropouts from this UVOT CDF-S data. We match the selected sources with available multiwavelength data from GOODS-South, MUSYC, and COMBO-17 to characterize the spectral energy distributions for these galaxies and determine stellar masses, star formation rates (SFRs), and dust attenuations. We compare these properties for LBGs selected in this paper versus $z\sim3$ LBGs and other CDF-S galaxies in the same redshift range ($0.5<z<2$), identified using photometric redshift techniques. The $z\sim1$ LBGs have stellar masses of $\langle \rm Log~M_*/M_\odot\rangle = 9.4\pm 0.6$, which is slightly lower than $z\sim3$ LBGs ($\langle \rm Log~M_*/M_\odot\rangle=10.2\pm 0.4$) and slightly higher compared to the $z\sim1$ CDF-S galaxies ($\langle \rm Log~M_*/M_\odot\rangle = 8.7 \pm 0.7$). Similarly, our sample of $z\sim1$ LBGs has SFRs (derived using both ultraviolet and infrared data, where available) of $\langle \rm Log~SFR/(M_\odot~yr^{-1})\rangle=0.7 \pm0.6 $, which is nearly an order of magnitude lower than $z\sim 3$ LBGs ($\langle \rm Log~SFR/{\rm M_\odot~yr^{-1}}\rangle=1.5 \pm 0.4$), but slightly higher than the comparison $z\sim1$ sample of CDF-S galaxies ($\langle \rm Log~SFR/{\rm M_\odot~yr^{-1}}\rangle=0.2 \pm 0.7$). We find that our $z\sim 1$ UV-dropouts have $\langle A_{\rm FUV}\rangle=2.0\pm 1.0$, which is higher than $z\sim3$ LBGs ($\langle \rm A_{\rm FUV}\rangle=1.0\pm 0.5$), but is similar to the distribution of dust attenuations in the other CDF-S galaxies ($\langle \rm A_{\rm FUV}\rangle\sim2.8\pm 1.5$). Using the GOODS-South multiwavelength catalog of galaxies, we simulate a larger and fainter sample of LBGs to compare their properties with those of the UVOT-selected LBG sample. We conclude that UVOT can be useful for finding and studying the bright end of 0.5$<z<$2.0 LBGs. \\ 
\end{abstract}

\section{Introduction}\label{sec:intro}
Initial interest in UV-selected galaxies began as an attempt to find the most primeval galaxies \citep[\eg those that were theoretically predicted by][]{PP1}. As young systems, such galaxies are expected to have recent star formation, low metal enrichment and little dust. In addition to bright UV continua, young, star-forming galaxies are expected to have strong breaks at 912 \AA, which occur as a result of the ground-state hydrogen ionization edge in the stellar absorption features of massive stars. The Lyman break technique exploits this feature in the rest-frame ultraviolet to isolate star-forming galaxies at great distances \citep{Steidel92, Steidel93, Steidel95, Steidel00}, and the ``UV-dropout'' products of this technique are the so-called Lyman break galaxies (LBGs). 

Despite having been selected as primordial systems, LBGs exhibit sufficient metal enrichment to exclude them from being the most primitive galaxies \citetext{\eg \citealp{pettini02}\ study the gravitationally lensed LBG, cB58, and find Type II supernovae residue, such as O, Mg, Si}. \cite{Mori} have conducted high resolution hydrodynamic simulations that follow the chemical evolution of primordial galaxies, finding that LBGs resemble infant versions of elliptical and bulge systems in the local universe. 

A decade of work has uncovered several significant results about LBGs at $z>2$ (see, e.g., \citealt{giarev}\ and references therein). LBGs form stars at intense rates, dominating the UV luminosity density at $z>2$. \cite{Bouwens, Bouwens08, Bouwens09, Bouwens2010} find only a modest decrease in the UV luminosity density out to $z=6$, indicating that LBGs represent a major phase in the early stages of galaxy formation and evolution. UV-selected galaxies are valuable for mapping the evolution of the global star formation rate density \citep{gia}, enriching the IGM \citep{Adel03}, locating large-scale structure and quantifying galaxy environments \citep{ouchi, adel05}. 

While studying UV-selected galaxies at $z>1$ is valuable, measuring their physical properties is challenging since high redshift studies are biased toward observing more luminous galaxies, and faint features are difficult to detect at these great distances. At $z<1$, \citet[further refined by \citealt{Choopes}]{Heckman05}, employ far-UV (FUV) luminosity and surface brightness criteria to select LBG-analogs, thereby named Lyman break analogs (LBAs). These $z\sim 0.2$ LBAs share several similar properties with LBGs: specific star formation rates (SFRs), metallicities, and attenuations \citep{Heckman05, Choopes, me2}.  When artificially redshifted to $z\sim3$, even their morphologies \citep{rod, Overzier2010} and ionized gas kinematics \citep{me-ifu, tsg} resemble those of actual LBGs. However, these galaxies are close enough to permit more detailed study of their physical properties. 

Combined, these separate UV-selected samples (LBGs and LBAs) provide insight into the evolution of this important galaxy population. Furthermore, they can provide pertinent information about the cosmic star formation history. At $z>$1, UV-selected galaxies are plentiful. They contribute significantly to the total UV luminosity density at those redshifts \citep{schimi04, Arnouts04}, but at $z<1$ these galaxies appear to be rare. Observations of galaxies at a multitude of wavelengths have shown that the star formation rate density (SFRD) of the Universe declines dramatically between $0 <z< 1$, peaking between $z\sim1-3$ \citep{Madau96, HB06}. The ``redshift desert'', named to signify the challenging nature of measuring redshifts for galaxies in the redshift regime $1<z<2$ \citep{Renzini2009}, contains valuable information that connects the peak of star formation with its rapid decline. It is likely that in this redshift range, the Hubble sequence took shape \citep{Papovich2005}. 

A few studies have used UV and optical data to expand the LBG selection to intermediate redshifts: $1<z<2$, connecting the low redshift ($z<1$) LBA sample with the high redshift ($z>2$) LBG sample. \cite{Ly2009} have targeted $1.5 <z<2.7$ LBGs, selected using the Galaxy Evolution Explorer \citep[\GALEX;][]{galex} near-UV (NUV) filter and deep Subaru optical filters; this work concludes that the peak of star formation occurred at $1.5<z<3$. \cite{Burgarella2006} have selected LBGs using the \GALEX~FUV$-$NUV color for galaxies in the COMBO-17 sample with known redshifts between $0.9 \lesssim z \lesssim 1.6$. This study compares infrared with ultraviolet observations and finds two populations: 24-micron detected and non-detected LBGs. The former can be classified as luminous infrared galaxies (LIRGs) with Log L$_{IR}>11$, exhibiting significant amounts of dust attenuation which is anti-correlated with the observed UV luminosity; the latter case appears to have little dust attenuation since stacking these LBGs constrains the infrared luminosity to be an order of magnitude less than the rest-frame NUV. Recently, two other studies have identified LBGs in this redshift range using Hubble Space Telescope (\HST) WFC3 data \citep{Hathi2010, Oesch2010} and have studied the evolution of luminosity function parameters with redshift. \cite{Oesch2010} find that the faint-end slope of the luminosity function, $\alpha$, appears to transition from a steep slope at $z>2$ to a flatter slope in the local universe. 

Since the 1$<z<2$ regime is a pivotal period in galaxy evolution, we ask: How do the properties of galaxies at these redshifts compare to local star-forming galaxies or to LBGs at high redshifts? Deep observations of some fields, such as the Chandra Deep Field South (CDF-S), have detected numerous non UV-selected galaxies at these redshifts allowing us to investigate: How do UV-dropouts compare to other galaxies of the same redshifts? 

In this paper we introduce another LBG sample that bridges the other aforementioned LBG samples across this relevant redshift range: 0.5 $\lesssim z \lesssim$ 2.  This sample of LBGs is selected using {\it Swift} \citep{Gehrels04} UV-Optical Telescope (UVOT) observations of the CDF-S. In this field, low redshift ($z \leq$ 2) and high redshift ($z \geq$ 3) LBGs have been identified \citep{Burgarella2006, Hathi2010, Oesch2010, Vanzella2009, gia2004, Bouwens, Bouwens08, Bouwens09, Bouwens2010}, yet none using UVOT. 

As a UV instrument, \Swift~UVOT offers a unique method for selecting LBGs, complementary to using \GALEX~and WFC3. \GALEX, with the largest  field-of-view (FOV; 1$^\circ$.2 diameter), covers the largest sky area of all these three UV instruments, yet is not as sensitive as WFC3 (\ie \GALEX~has imaged $\sim 4700 \deg^2$ down to 22.5 magnitudes as of December 2010\footnote{Information available on \GALEX~Legacy Survey website: {\texttt http://galexgi.gsfc.nasa.gov/docs/galex/Documents/GALEX-Legacy-Survey.html}} and 80~deg$^{2}$ to 25.0 magnitudes in the Deep Imaging Survey). However, UVOT offers two advantages over \GALEX: multiple NUV filters (\GALEX~only has one wide NUV filter), which allows the selection of UV-dropout candidates based on UVOT data alone, and better spatial resolution (FWHM$\approx$2\arcsec.7, compared to FWHM$\approx$5\arcsec~for \GALEX). WFC3 has excellent spatial resolution (0\arcsec.2) and sensitivity ($\geq$500 times that of UVOT; see Figure \ref{fig:filters}); it also has multiple UV filters for UV-dropout selection. However, the UVOT FOV (17\arcmin$\times$17\arcmin) is significantly larger than WFC3's FOV (2.7\arcmin$\times$2.7\arcmin) and \Swift~observations have covered a large area of the sky (\eg there are 40 GRB fields with $\approx$200 ks of exposure time in the UV filters). We can use UVOT observations to select these rare objects to study the bright end of the LBG population. Here we discuss the utility of using \Swift~UVOT to select LBGs from the ``redshift desert''. 

The CDF-S region of the sky has been extensively and deeply covered by multiwavelength observations. We use available datasets in several ways: to constrain various properties of the UVOT-selected LBGs; to provide a large comparison sample of other photometrically determined $0.5<z<2.0$ galaxies; and to draw from this large parent population a sample of simulated UV-dropout galaxies to compare their properties with the observed UVOT-selected LBGs. In Section \ref{sec:data}, we present the sample selection and data analysis. We discuss the LBG candidates in Section \ref{sec:results} and present our results for the $0.5<z<2.0$ LBG sample and compare these LBGs with the other $0.5<z<2.0$ galaxies (including a simulated LBG sample, defined in Section \ref{sec:data}) and other UV-selected populations (LBAs and $z\sim$3 LBGs). In Section \ref{sec:end} we summarize our main results and discuss the merit of studying intermediate redshift ($0.5<z<2.0$) LBGs with \Swift. Throughout our analysis, we use the following cosmology: H$_0$=70 km s$^{-1}$ Mpc $^{-1}$, $\Omega_m$=0.30, and $\Omega_\Lambda$= 0.70, and assume a \cite{Kroupa} initial mass function (IMF). 

\section{Data Analysis}\label{sec:data}
\subsection{UVOT Data}\label{sec:uvot}
While the primary mission of the \Swift~satellite is to study gamma-ray bursts (GRBs), the UVOT instrument has served in the study of  
supernovae, galaxies, and active galactic nuclei (AGN), amongst other subjects.  The UVOT is a 30 cm telescope with f-ratio 12.7 \citep{Roming05}.  The CDF-S was observed by UVOT between July 2007 and December 2007 for $\sim$450ks.

\begin{figure}[t!]
\begin{center}
  \includegraphics[width=3.5in]{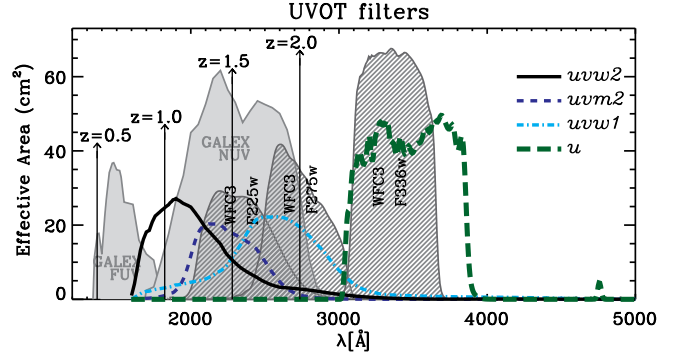}
\caption{We compare {\it Swift}~UVOT filter curves ($uvw2$ in solid black, $uvm2$ in dashed blue, $uvw1$ in dash-dotted cyan and $u$ in thick dashed dark green) to other UV filter curves--  \GALEX~FUV and NUV (marked and shaded light gray), and WFC3 UV filters (F225W, F275W and F336W; marked and shaded in dark gray, scaled by 1/500 to fit). The location of the Lyman break feature is shown for $z=$0.5, 1.0, 1.5 and 2.0 for reference. The $uvw2$ and $uvw1$ filters suffer from a ``red tail'', with a shallow decrease at longer wavelengths.}\label{fig:filters}
\vspace{-0.1in}
\end{center}
\end{figure}

The UVOT instrument has two grisms and seven broadband filters. \cite{Poole08} and \cite{Breeveld10} provide detailed discussion about the UVOT filters and detectors. In our analysis, we focus mainly on the ultraviolet filters: $uvw2$ ($\lambda_c=$1928 \AA; FWHM$=$657 \AA), $uvm2$ ($\lambda_c=$2246 \AA; FWHM$=$498 \AA), $uvw1$ ($\lambda_c=$2600 \AA; FWHM$=$693 \AA) and the $u$ ($\lambda_c=$3465 \AA; FWHM$=$785 \AA) filter. Figure \ref{fig:filters} shows these filter curves and compares them to \GALEX~and \HST~WFC3 UV filters. 

\begin{figure}
\begin{center}
  \includegraphics[width=3.5in]{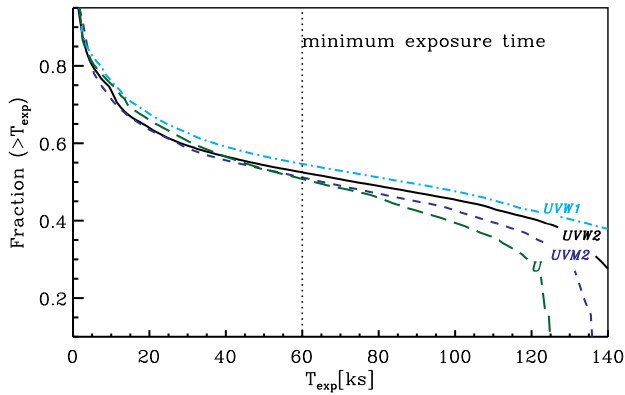}
    \end{center}
  \caption{The exposure times across the field vary smoothly for all filters: $uvw2$ (solid black), $uvm2$ (dashed blue), $uvw1$ (dash-dotted cyan) and $u$ (thick dashed dark green) between exposure times, T$_{\rm exp}=$40$-$100ks. Within this range of exposure times, going deeper (to longer exposure times) does not sacrifice field coverage, while it does increase signal-to-noise. To ensure 5$\sigma$ detections, our selection criterion requires $u<$24.5 and exposure times $>$60ks. More than 50\% of the field has been observed longer than this minimum exposure time, 60ks (marked by dotted line), in all filters. }
   \label{fig:fieldhist}
\vspace{0.05in}
   \end{figure}

The CDF-S was observed during several separate observations, and the data were combined using standard \Swift~packages\footnote{\texttt{http://heasarc.gsfc.nasa.bov/docs/swift/analysis}}. The maximum exposure times per filter range from 125$-$145 ks, depending on the filter \citep[see][for a thorough discussion of the \Swift~CDF-S observations]{Hoversten10}. Figure \ref{fig:fieldhist} shows the fraction of the field that was observed longer than some exposure time. In this figure, we show that the exposure times across the field vary smoothly for all filters: $uvw2$ (solid red), $uvm2$ (dashed black), $uvw1$ (dash-dotted blue) and $u$ (thick dashed dark green) between exposure times, T$_{\rm exp}=$40$-$100ks.  

A catalog of sources was generated following \citet{Hoversten10} with some modifications.  A summary of this process with the differences highlighted is as follows.  UVOT observations have gone through an initial processing by the UVOT pipeline\altaffilmark{6}.  This pipeline provides corrected image files and exposure maps for each observation.  The fine aspect correction applied to images is now also applied to the exposure maps.  However this was not the case at the time of the CDF-S observations.  Our CDF-S observations were processed by the Swift Data Center using the latest version of the pipeline (version 2.2), but this is not yet available in the {\it Swift} archive.  Alternatively, one can create fine aspect corrected exposure maps from the images, UVOT housekeeping data, and spacecraft auxiliary files available in the archive as described in \citet{Hoversten11}.

Images and exposure maps were summed using the publicly available UVOT FTOOLS (HEAsoft 6.6.1)\footnote{\texttt{http://heasarc.gsfc.nasa.gov/docs/software/lheasoft/}}.  This involved two flux conserving interpolations of the images.  A correction for known bad pixels was applied in the UVOT pipeline, and cosmic ray corrections were not necessary for UVOT images.  Images were divided by the relevant exposure map to generate count rate images.

The count rate images were analyzed using SExtractor \citep{sex}. The list of SExtractor parameters used can be found in Table 2 of \citet{Hoversten10}.  There are several types of magnitudes which are calculated by SExtractor. The {\tt MAG\_AUTO} magnitudes are designed to measure the total magnitudes of the galaxies. However, for LBG selection it is most important to have accurate colors for the galaxies.  For this reason the {\tt MAG\_ISO} isophotal magnitudes, which are the recommended magnitudes for studying colors, were used.  Apertures were determined from the $u$ band image and then the same apertures were applied to all four images.  The $u$ band was used for aperture determination because candidate LBGs are expected to drop out of the bluer filters, and would likely be missed if the apertures were determined from the NUV filters.

The segment map, output by SExtractor, containing the apertures of sources was cleaned using the Markov chain algorithm described in Appendix B of \citet{Hoversten11}.  Then the IDL code described in \citet{Hoversten11} was used to convert SExtractor count rates to magnitudes. This code was created to take a user specified segment map, and additionally to apply UVOT corrections not included within SExtractor.  One of these is the ``coincidence-loss'' correction described in \citet{Poole08}.  As UVOT is a photon counting detector it suffers from an undercounting of photons which becomes progressively worse, and eventually uncorrectable, for brighter sources.  At the faint magnitudes of the CDF-S this correction is negligible; only one source in the CDF-S has a correction of more than 1\% and even then it is only this large in the $u$ filter.  Secondly this code applies the updated UVOT zeropoints (found in the UVOT Digest webpage: {\texttt{http://swift.gsfc.nasa.gov/docs/swift/analysis/ uvot\_digest/zeropts.html}}) and the AB magnitude correction \citep{Breeveld11} to put the UVOT magnitudes on the AB system.

To summarize the differences between the \citet{Hoversten10} photometry and that used here, isophotal magnitudes are used instead of {\tt MAG\_AUTO} ``automatic aperture'' magnitudes, apertures are defined uniformly in the $u$ band rather than on a filter by filter basis, a coincidence loss correction is applied (although negligible), and updated UVOT zero points and AB corrections are applied. 

\begin{figure*}
\begin{center}
\includegraphics[width=7.0in]{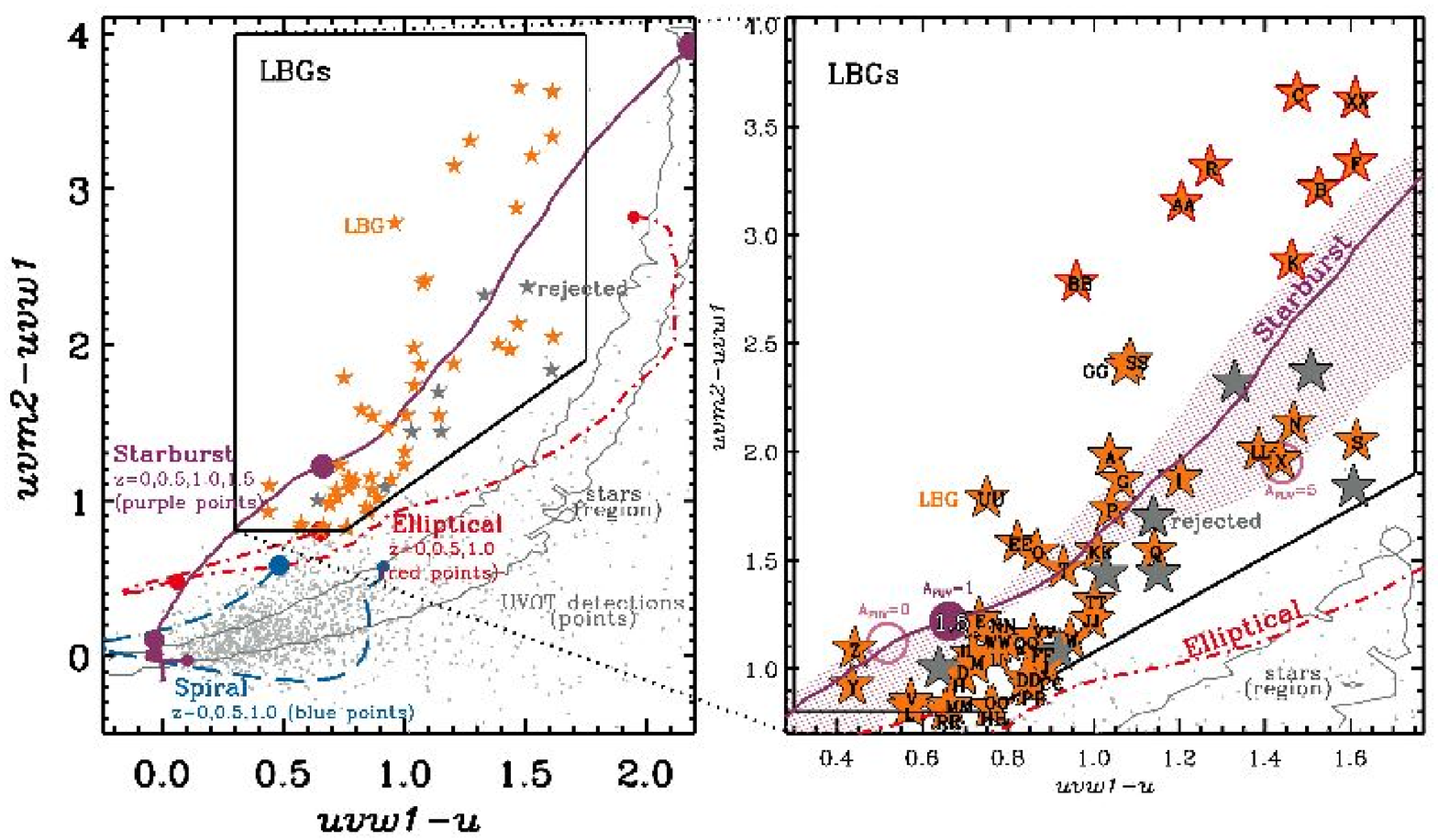}
\caption{We select LBG candidates based on $uvw1-u$ and $uvm2-uvw1$ colors (the selection region is outlined in black based on Equations \ref{eqn:lbg2}$-$\ref{eqn:lbg4}, with LBGs shown as orange stars and rejected candidates shown as dark gray stars). To compare with other galaxy populations, we show gray points to mark other objects from the CDF-S UVOT catalog, and mark evolution tracks for three types of galaxies (early type elliptical galaxies, late type spirals and starbursts shown with dash-dotted red, long-dashed blue and solid purple, respectively). The dark gray outline shows the region populated by stars, using the \cite{ckstars} model spectra to span a large range of stellar temperatures, surface gravities, metallicities, and alpha enhancements. The tracks include points to mark every $\Delta z=$0.5 with increasing point size. The right panel zooms into the LBG-selected region, labeling every LBG with a letter (`A' thru `ZZ'; see Section \ref{sec:results}); the purple point on the starburst track marks $z=1.5$, as labeled. The shaded region shows the effects of dust attenuation on the starburst track, ranging from A$_{\rm FUV}=0$ (top edge) to 5.0 (bottom edge), with light purple open circles marking $z=1.5$; the solid purple starburst track assumes A$_{\rm FUV}=1.0$. LBGs outlined in red have photometric errors $\sigma(uvm2)>1$ mag, potentially accounting for their deviation from the starburst model track. }\label{fig:sel}
\vspace{0.08in}
\end{center}
\end{figure*}

\subsection{Selection of LBG Candidates}
In Figure \ref{fig:fieldhist}, we show that for exposure times between 40$-$100 ks, going deeper (to longer exposure times) does not sacrifice field coverage, while it does increase signal-to-noise. From the UVOT CDF-S catalog, we selected only sources with exposure times exceeding 60ks in all filters (shown as a dotted line in Figure \ref{fig:fieldhist}) and $u<24.5$ mag in order to ensure reliable detections (signal-to-noise in $u$ filter exceeding 5$\sigma$). 


We find that in regions that were observed for at least 60ks, we are $\sim 29$\% complete at our limiting magnitude ($u=24.5$) with completeness increasing for brighter magnitudes and in regions with longer exposure times (\eg completeness is $\sim80$\% at $u=23$ for 60ks and $\sim 47$\% at $u=24.5$ for 122ks; see \citealt{Hoversten10}\ for more details regarding the completeness calculation). We note that this paper focuses on the \Swift~UVOT selection of potential LBG candidates and their properties, which does not rely upon completeness. 

The $0.5<z<2$ UV-dropout selection takes into account the unique filter curves of the UVOT instrument (see Figure \ref{fig:filters}). The extended red tails in the $uvw2$ and $uvw1$ filter curves are discussed in more detail in \cite{Brown10}. The color selection requires observations in a minimum of three filters. Observing in three filters, rather than applying the dropout criterion based on two filters, prevents selecting interlopers, such as red stars and red $z\sim$0 galaxies.
   
While $uvw2$ is the bluest UVOT filter, the red tail prevents a clean separation of the star-forming galaxies from other types of sources. Rather, we find that the $uvm2$, with its relatively steep edges works nicely as the bluest filter for UV-dropouts. As Figure \ref{fig:filters} displays, the Lyman break feature causes objects at $z=1.5$ to drop out from $uvm2$. 

In the left panel of Figure \ref{fig:sel} we show where the objects found in the UVOT catalog (described in \S\ref{sec:uvot}; gray points) are located in $uvw1-u$ versus $uvm2-uvw1$ color space, along with evolutionary tracks of three types of galaxies (early type elliptical galaxies, late type spirals, and starbursts shown with dash-dotted red, long-dashed blue and solid purple, respectively), with points marking $z=0$, 0.5, 1.0 with increasing size on all the tracks and additionally $z=1.5$ on the solid purple starburst track. 

The galaxy tracks were calculated using P\'EGASE \citep{peg} galaxy models, redshifted and convolved with the UVOT filters to get magnitudes at each redshift. The elliptical galaxy model assumes a simple stellar population (SSP) model with solar metallicity at an age  of 12 Gyr with A$_{\rm V}=$1.0 with Milky Way extinction from \cite{Pei92}. The spiral galaxy model assumes a constant star formation history (SFH) with solar metallicity at an age of 12 Gyr with A$_{\rm V}=$1.0 also with Milky Way extinction. The starburst model assumes a constant SFH with solar metallicity at an age of 100 Myr with A$_{\rm V}$=0.4 following the \cite{Calzetti94} dust law (\ie A$_{\rm FUV}= 2.48 \times {\rm A}_{\rm V} \approx $1.0); in addition, the shaded region (on right panel) shows a range of dust attenuation values (0 $<A_{\rm FUV}<5$, top edge of region to bottom edge), with $z=1.5$ marked with light purple open circles. The starburst model track includes the Lyman alpha forest derived in \cite{Madau95}, which becomes more important with increasing redshift. 

Our selection region aims to avoid the region occupied by stars (shown by the dark gray outline). This region was calculated using a grid of model stellar spectra from  \cite{ckstars}, spanning a range of effective temperatures ($3500 {\rm K} < {\rm T_{eff}} < 50000$ K), surface gravities (0.0 $< \log {\rm g} <5.0$), and metallicities (-4.0 $< [M/H] < 0.5$, where [M/H] is the ratio of metals to hydrogen) and two values of alpha-enhancements ($[\alpha/{\rm Fe}]=$0.0 and 0.4, where [$\alpha$/Fe] is the ratio of $\alpha$-elements, \ie O, Mg, Si, Ca, Ti, to iron).

We used the COMBO-17 photometric catalog to get preliminary redshifts for all the galaxies with $uvm2-uvw1>0.5$ and $0<uvw1-u<2$ and then determined the boundaries of the selection region based on maximizing the number of galaxies with redshifts $z>0.5$. For example, the slope and lower limit in $uvm2-uvw1$ were set by minimizing the number of $z<0.5$ contaminating sources (stars and low redshift elliptical galaxies).  Based on this exercise, we determined that the following equations best separate intermediate redshift LBG candidates from other populations: 

\begin{align}
20.75 {\rm ~ mag} < u < 24.5 {\rm ~ mag}  ~~ &\wedge \label{eqn:lbg1}\\ 
0.3 {\rm ~ mag} < uvw1 -u < 1.75 {\rm ~ mag}~~  &\wedge \label{eqn:lbg2}\\
0.8{\rm ~mag} < uvm2- uvw1 <  4  {\rm ~ mag}~~ &\wedge \label{eqn:lbg3} \\
1.1(uvw1 - u) - 0.025 {\rm ~ mag} < (uvm2-uvw1)\label{eqn:lbg4}
\end{align}

In Figure \ref{fig:sel}, the black outline encloses the LBG-selection region described by Equations \ref{eqn:lbg2}$-$\ref{eqn:lbg4}. Applying this selection, we select 58 LBG candidates; 50 of these candidates have photometric ($\langle \Delta z/(1+z)\rangle \sim 0.1$ for these galaxies) or spectroscopic redshifts consistent with our selection ($0.5<z<2.0$; shown as orange stars) while the remaining 8 candidates were rejected because of inconsistent or questionable redshifts or photometry (described in more detail in \S\ref{sec:res_cand}; shown as gray stars). The sources outlined in red have large UVOT photometric errors (specifically, $\sigma(uvm2) > 1$ mag); generally, these are also the objects farthest from the starburst model track. The medians of other UVOT photometric errors are: $\sigma(uvm2)$ , $\sigma(uvw1)$, and $\sigma(u)=0.5$, 0.2, and 0.1 mags.  

The observed UVOT magnitudes and relative exposures (exposure time compared to the maximum exposure for the field) for the candidates are shown in Table \ref{tab:raw}; the photometric (or spectroscopic, where available) redshifts for the candidates are shown in the fifth column of Table \ref{tab:derived} along with other derived properties based on multiwavelength data, which are discussed in the following sections. 

A magnified view of the selection region is shown in the right panel of Figure \ref{fig:sel}. Note that the LBG candidates are selected to lie close to the starburst curve between $0.5<z<2$ (the marked and labeled purple point on the starburst track shows $z=1.5$ in this panel). Candidates are labeled by letter (likely LBGs) or identification name (rejected candidates)-- we discuss the characteristics of all the candidates in Section \ref{sec:results}. To differentiate between the LBG sample and the rejected candidates throughout this paper, we label the LBGs as A through XX, arbitrarily named by their position (ordered clock-wise) in Figure \ref{fig:cdfs-lbgs}. The thumbnails surrounding the UVOT CDF-S image show enlarged views for the LBGs -- each thumbnail is 20\arcsec~per side (corresponding to a physical scale of $\sim 170$ kpc at $z=$1.5), and the orange circles have diameters of 2.7\arcsec, which roughly corresponds to the UVOT point spread function (PSF). 

\subsection{Multiwavelength Data and Broadband SED Fitting} \label{sec:opt}
\begin{figure}
\includegraphics[width=3.5in]{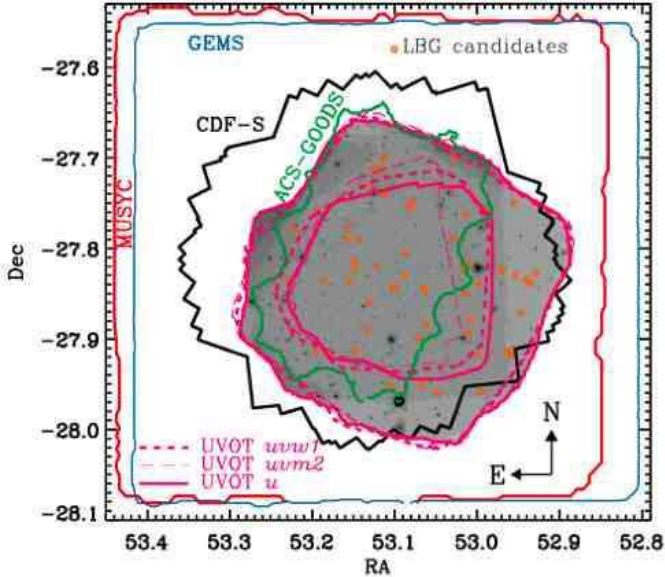}
\caption{Schematic of CDF-S coverage by various multiwavelength programs: COMBO-17 (encompassing this entire field and too large to display outlines), MUSYC (red), \HST~GEMS (cyan), $Chandra$ (black), \HST~ACS-GOODS (green), and \Swift~UVOT (magenta; thick dashed, thin dashed and solid lines mark $uvw1$, $uvm2$ and $u$ filters, inner contours refer to regions within 98\% of maximum exposure time). Background image is UVOT $u$. LBG candidates are marked with orange circles. }\label{fig:field}
\vspace{0.1in}
\end{figure}

To study these galaxies in detail, we benefit from the availability of rich multi-wavelength data in this field. Figure \ref{fig:field} shows the UVOT image compared to several other well-studied programs for the CDF-S and extended CDF-S. We match our candidates with sources in the K-band selected Multiwavelength Survey by Yale-Chile \citep[MUSYC;][]{musycK}, Advanced Camera for Surveys (ACS) images in the Great Observatories Origins Deep Survey(GOODS) South \citep[ACS-GOODS; ][]{Dahlen2010}, and COMBO-17 \citep{combo17, Wolf2008} datasets. These catalogs contain photometric data covering the optical, near-infrared wavelengths with several filters (MUSYC, ACS-GOODS and COMBO-17 have 10, 7 and 17 filters, respectively). 

The UVOT imaging resolution is $\approx$ 2.7\arcsec. For proper matching between the 58 UVOT-selected LBG candidates (including the rejected candidates) and these other catalogs, we first compared the astrometry of $>5000$ sources in the UVOT catalog with brightest sources within 2\arcsec~from the COMBO-17 catalog (which was found to match MUSYC and ACS-GOODS astrometry). We find that the UVOT positions are slightly offset: RA(COMBO-17) $=$ RA(UVOT)$+$0\arcsec.9 ($\pm 0\arcsec.4$) and Dec(COMBO-17) $=$ Dec(UVOT)$-$0\arcsec.5 ($\pm 0\arcsec.4$). After applying these corrections, the UVOT candidates were matched to sources in the other catalogs within 2\arcsec, which allows matching a COMBO-17 (FWHM$=$1\arcsec.5) source with a UVOT (FWHM$=$2\arcsec.7) source. Only 6 cases were matched to a single source (S, T, X, DD, HH, and J033145.7$-$275003.5). In cases of multiple matches, the brightest R-band (or $z_{850}$, for ACS-GOODS) source was selected. J033207.4$-$274400.4 was eliminated from our analysis since it appeared to have no optical match and the UVOT image showed a possible artifact from a nearby bright object.

We tested the likelihood for multiple matches between any UVOT catalog source and a source in COMBO-17 or ACS-GOODS. We find that multiple sources are matched $\sim30$\% of the time. Therefore, it is significant that only 6 sources (or 10\%) of LBG cases have matches to single sources and may suggest that these are galaxies found in pairs or group environments, as seen for $z\sim1$ LBAs by \cite{me-env}, and may be experiencing triggered star formation. 

Combining the UVOT photometry with matched photometry from these other catalogs poses some challenges, since the techniques used to determine the photometry in each case vary. For MUSYC photometry, the ``color flux'' in each filter is determined by SExtractor's {\tt MAG\_ISO}, which measures the flux within the isophote (set to be at least 2\arcsec.5) corresponding to the lowest detection threshold. This value is corrected to the total flux by applying a correction factor calculated based on the ratio of total flux in the K-band (using SExtractor's {\tt MAG\_AUTO}, which accounts more fully for the size and shape of the light distribution) to the color flux in the K-band. COMBO-17 has a similar method, also correcting the color flux of each filter into a total flux by using the R-band to scale the R color flux into the total flux in R. However,  they do not use SExtractor for the color fluxes, but another package which uses seeing-adaptive, weighted aperture photometry to equalize the effects of seeing on data from different bands; they use SExtractor's {\tt MAG\_BEST}, which is similar to {\tt MAG\_AUTO} but also corrects for contamination from neighboring sources, to determine the total flux \citep[for more details, see][]{combo17}. The ACS-GOODS photometry uses a template fitting technique which matches the high-resolution ($\sim$0\arcsec.1) $z$-band image to the lower resolution ($\sim$2\arcsec) infrared (IR) images to reduce blending effects and provide consistent photometry in all bands regardless of the PSF size \citep{Dahlen2010}. 

\cite{musycK} compares MUSYC photometric offsets with different datasets, including COMBO-17 and ACS data, and with stellar SEDs. We note that the ACS data compared in their work is from FIREWORKS \citep{FIREWORKS}, which is K$_s$-selected, whereas our ACS-GOODS data is z-selected -- however, this distinction has negligible effect for comparing the photometry from these datasets. Our own comparisons of the photometry for matched sources between the COMBO-17, ACS-GOODS and MUSYC datasets agree with \cite{musycK}. For our sample, we analyze all these datasets together in order to provide the most complete information about their SEDs. However, reconciling differences between photometry from different datasets can add complexity to interpreting the cause of those inconsistencies. We point out specific examples of this in \S\ref{sec:sed}.

We use {\it kcorrect} \citep[version 4.2;][]{kcorrect2} to perform SED-fitting, k-corrections, and estimate masses. The {\it kcorrect} program uses the non-negative matrix factorization technique to fit a set of basis template models to the photometry; the full set of templates include 450 instantaneous burst templates from \cite{BC03} stellar synthesis models and 35 emission-line templates for ionized gas from MAPPINGS-III \citep{Kewley01}. \cite{kcorrect2} have determined that the linear combination of 5 basis templates are sufficient to describe the spectra of most galaxies.

In our sample, 25 candidates have spectroscopic redshifts and all have photometric redshifts (except J033207.4$-$274400.4 has no optical match, thus no known redshift). Based on comparing the spectroscopic redshifts (where available) to the photometric redshifts from the other datasets, we find that COMBO-17 gives the most reliable redshifts. Therefore we use the spectroscopic redshifts when available and COMBO-17 redshifts \citep{Wolf2008} in other cases. However, in six cases (LBGs C, K, O, SS, UU, and XX) the COMBO-17 redshifts were too low and did not fit the SED well. The photometric redshift from MUSYC was used for LBG O and the photometric redshifts from GOODS \citep{Dahlen2010} were used for the other 5 galaxies, since we found these catalogs provided the best redshifts for these galaxies. The redshifts used for SED-fitting and their source are listed in Table \ref{tab:derived}, as well as the estimated stellar masses and rest-frame UV SFRs.  

\subsection{Our LBG Candidates}\label{sec:res_cand}
\begin{figure*}
\begin{center}
 \includegraphics[width=6.in]{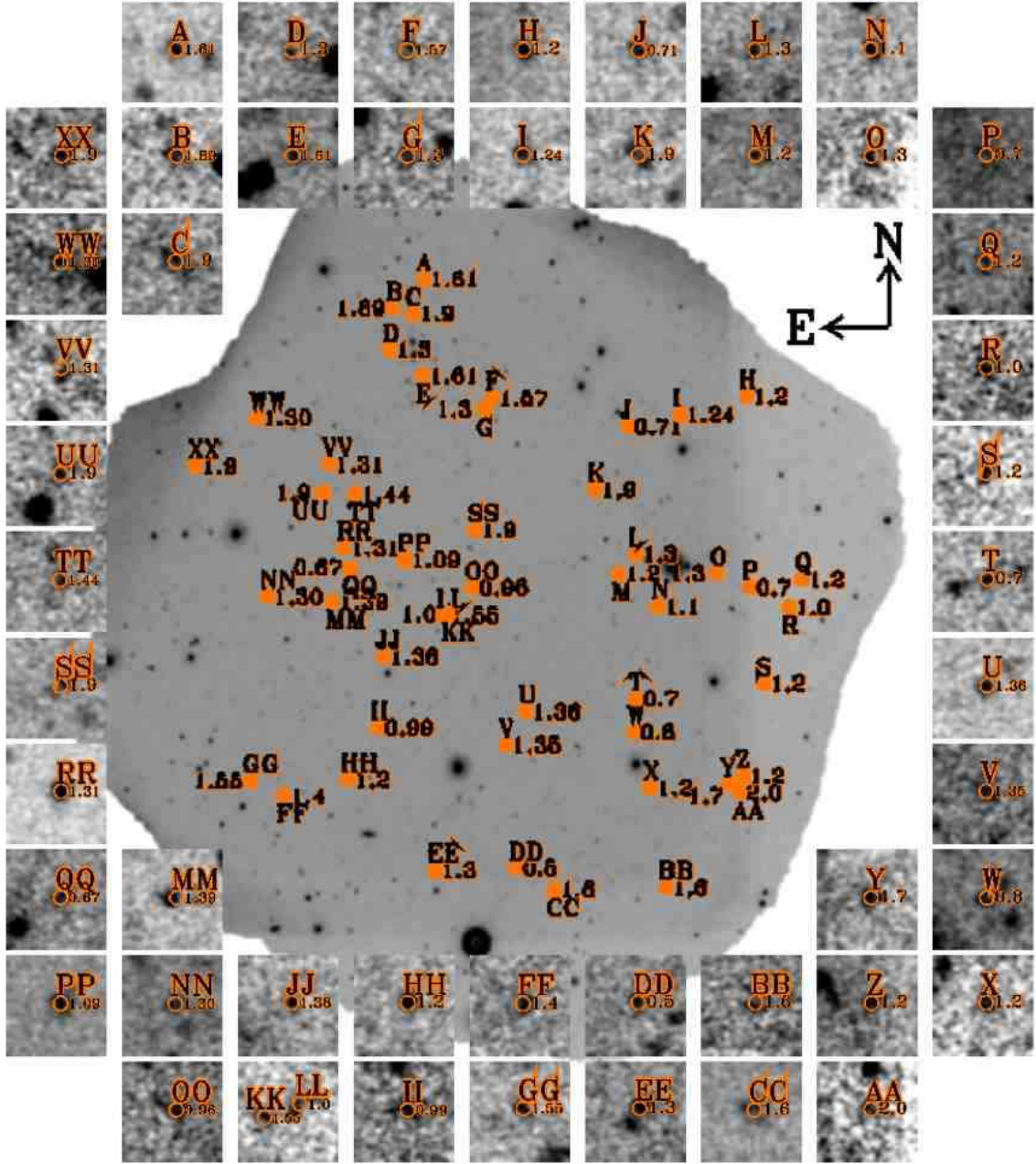}
\caption{{\it Swift} UVOT $u$ image-- 
LBGs are marked on the center CDF-S field image as orange points, and labeled A-XX, with corresponding redshifts labeled. Zoomed-in views of the UVOT images for each LBG is shown in the surrounding panels, 20\arcsec~ per side (corresponding to $\sim170$kpc at $z=1.5$)-- the orange circles have 2.7\arcsec~diameters, roughly the UVOT PSF FWHM. }\label{fig:cdfs-lbgs}
\vspace{0.05in}
\end{center}
\end{figure*}

\begin{figure*}[h!]
\begin{center}
  \includegraphics[width=6.0in]{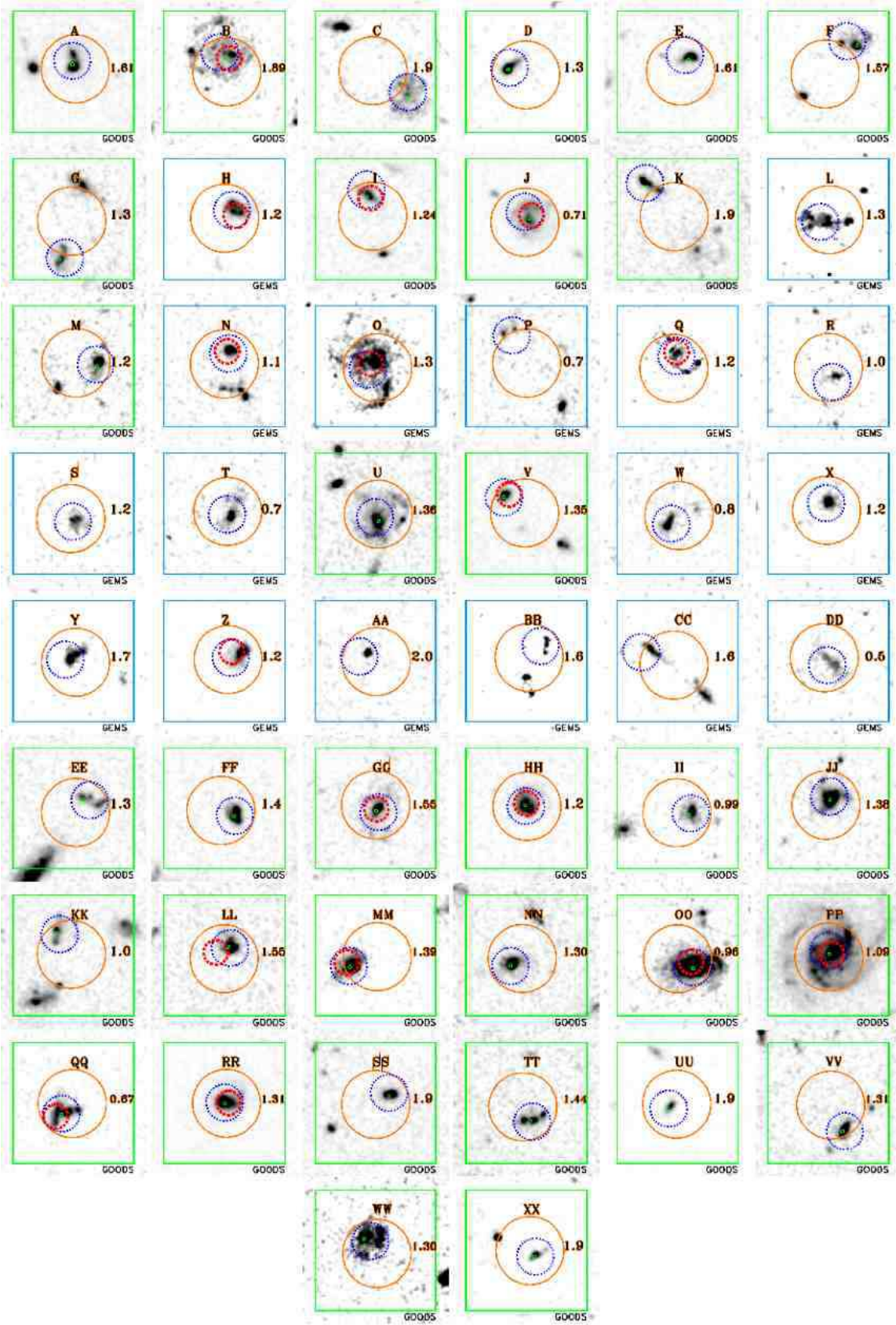}
  \end{center}
  \vspace{-0.1in}
  \caption{Optical z-band images of $z=$0.5$-$2 LBGs -- \HST/ACS data comes from the GOODS-S or GEMS datasets, labeled on lower right corner and marked with green or cyan (respectively) boxes sized to 5\arcsec~on a side (at $z=1.5$, $\sim 43$kpc). All images are oriented North up and East to the left. LBGs are labeled as in Fig. \ref{fig:cdfs-lbgs}, with corresponding redshifts (either photometrically or spectroscopically determined, see Table \ref{tab:derived}) labeled on the right of the image. Circles mark PSFs for the source detections in UVOT (solid orange; 2\arcsec.7 FWHM), MUSYC (thick dashed red; 1\arcsec~FWHM), ACS/GOODS (solid green; 0\arcsec.11 FWHM) and COMBO-17 (thin dotted blue; 1\arcsec.5 FWHM) catalogs. }
   \label{fig:mapsbw}
   \end{figure*}

We use the redshifts derived from other catalogs (see Table \ref{tab:derived}) either spectroscopically or photometrically to determine whether the LBG selection identified high redshift (0.5$<z<$2.0) galaxies rather than low redshift interlopers or artifacts. Out of the 58 candidates, 50 are satisfactory. These 50 LBGs are marked with their corresponding redshifts in Figure \ref{fig:cdfs-lbgs}. Figure \ref{fig:mapsbw} shows the \HST/ACS optical images for these LBGs. These images come from the GOODS-S and GEMS datasets. The GOODS-S and GEMS z$_{850}$-band images ACS images are displayed. The 2.7\arcsec~diameter circles marking the UVOT PSF appear in orange, with larger 5\arcsec~squares in green for GOODS-S (33 sources) and in cyan for GEMS (17 sources). The smaller apertures correspond to the MUSYC (thick dashed red; 1\arcsec.0 FWHM), COMBO-17 (dotted blue; 1\arcsec.5  FWHM), and ACS (solid green; 0\arcsec.11 FWHM) PSFs. 

\begin{figure*}
\begin{center}
  
 
 \includegraphics[width=4.0in]{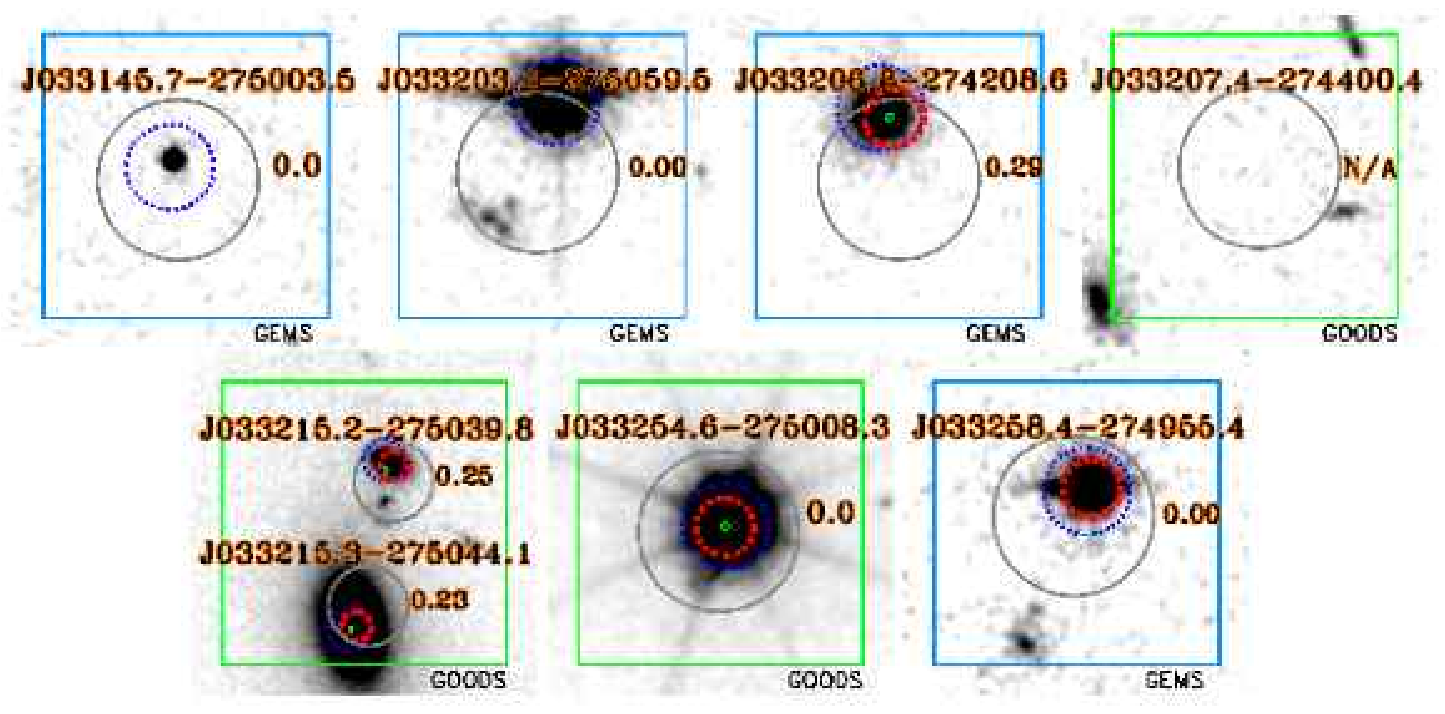}
  \end{center}
  \caption{Optical thumbnails of rejected candidates. Same as Figure \ref{fig:mapsbw} except that the UVOT apertures marking failed LBGs are gray circles and labeled by their ids. The field of view for objects J033215.2$-$275039.8 and J033215.3$-$275044.1 (both shown together in the fourth panel) is twice that of the other images -- therefore, the green box surrounding these two objects is 10\arcsec~per side (rather than 5\arcsec).}
   \label{fig:mapsrejbw}
\vspace{0.05in}
   \end{figure*}

The remaining 8 objects are eliminated from any further analysis, and briefly discussed below. We display the images for the rejected candidates in Figure \ref{fig:mapsrejbw}. In five cases (J033203.4$-$275059.5, J033207.4$-$274400.4, J033215.3$-$275043.7, J033254.6$-$275008.3, and J033258.4$-$274955.4) visual inspection of the optical images confirm our choice to reject candidates with photometric redshift $z<0.5$; however in the other cases, this choice is not so obvious -- we have elected to discard sources with confusing or ambiguous matches. 

\noindent$\bullet$~J033145.7$-$275003.5 -- The photometric redshift for this object in COMBO-17 is $z=0.03$, and the object is not found in any other catalog. While it is possible that the photometric redshift is incorrect and that this object may be an LBG, we choose to remove it from further analysis. \\
\noindent$\bullet$~J033203.4$-$275059.5 -- The nearest and brightest optical match is a star (appearing with diffraction spikes in Figure \ref{fig:mapsrejbw}). While a fainter source to the southeast has been identified as a galaxy in the 0.5$<z<$2 range, contamination from the bright star makes the detection and analysis of this source unreliable. Therefore we reject this object from our analysis.\\  
\noindent$\bullet$~J033206.8$-$274208.6 -- The optical match for this source appears to be an elliptical galaxy, with a spectroscopic redshift of $z=0.29$. Since, this object is unlikely to be an LBG, we remove it from our sample.\\
\noindent$\bullet$~J033207.4$-$274400.4 -- The UVOT image shows a bright object in the southeast and the detected ``source'' appears to be an artifact from this brighter object. The optical image shows no obvious counterpart. We reject this source based on the questionable UVOT detection. \\
\noindent$\bullet$~J033215.2$-$275039.8 -- The bright optical counterpart appears to be a red, elliptical galaxy at $z=0.2$. There is a very blue, fainter object to the south (within the UVOT 2.7\arcsec~FWHM PSF). But we exclude this object because we find no optical match for the fainter, blue object in COMBO-17, ACS-GOODS, or MUSYC. \\
\noindent$\bullet$~J033215.3$-$275044.1 -- The optical image for this source clearly displays a red, elliptical galaxy. With a spectroscopic redshift of $z=0.23$, this object is unlikely to be an LBG and is removed from our sample. \\
\noindent$\bullet$~J033254.6$-$275008.3 -- The optical match is a very bright object (appearing with diffraction spikes in the image, see Figure \ref{fig:mapsrejbw}). Therefore we reject this object from our analysis.  \\
\noindent$\bullet$~J033258.4$-$274955.4 -- The only obvious source in the optical image is a bright star. Therefore we eliminate this object from our analysis.  \\

\begin{deluxetable*}{clllccccrrrr}
\tabletypesize{\scriptsize}
\setlength{\tabcolsep}{0.07in}
\tablecolumns{12} 
\tablewidth{0pc} 
\tablecaption{Summary of UVOT Photometry for LBG Candidates} 
\tablehead{
\colhead{ID} & \colhead{J2000 ID} & \colhead{RA} & \colhead{Dec} & \colhead{$uvw2$} & \colhead{$uvm2$} & \colhead{$uvw1$} & \colhead{$u$} & \colhead{T\tablenotemark{*}($uvw2$)} &\colhead{T\tablenotemark{*}($uvm2$)} & \colhead{T\tablenotemark{*}($uvw1$)} & \colhead{T\tablenotemark{*}($u$)} \\
& & & & \colhead{(mag)} &  \colhead{(mag)}  &  \colhead{(mag)} &  \colhead{(mag)} & \colhead{(\%)} & \colhead{(\%)} & \colhead{(\%)} &  \colhead{(\%)}
}
\startdata
\cutinhead{LBG Sample}
A & J033226.8$-$274156.6 &  53.1117  &  $-$27.6991 &   25.4 &   26.0 &   24.0 &   23.0  &   94.5  &   89.6  &   79.1  &   82.5   \\
B & J033230.3$-$274241.0 &  53.1264  &  $-$27.7114 &   28.2 &   28.8 &   25.6 &   24.0  &   96.0  &   93.7  &   83.7  &   87.1   \\
C & J033228.0$-$274249.9 &  53.1167  &  $-$27.7139 &   26.5 &   28.6 &   25.0 &   23.5  &   96.1  &   97.6  &   85.8  &   89.2   \\
D & J033230.6$-$274345.6 &  53.1279  &  $-$27.7293 &   26.6 &   26.0 &   25.0 &   24.3  &  100.0  &  100.0  &   98.4  &   94.3   \\
E & J033226.8$-$274425.0 &  53.1119  &  $-$27.7403 &   25.8 &   25.9 &   24.6 &   23.9  &  100.0  &  100.0  &  100.0  &  100.0   \\
F & J033218.8$-$274500.0 &  53.0784  &  $-$27.7500 &   27.2 &   29.1 &   25.7 &   24.1  &  100.0  &  100.0  &  100.0  &  100.0   \\
G & J033219.8$-$274516.2 &  53.0827  &  $-$27.7545 &   27.1 &   27.3 &   25.4 &   24.4  &  100.0  &  100.0  &  100.0  &  100.0   \\
H & J033149.4$-$274456.3 &  52.9560  &  $-$27.7490 &   25.6 &   24.7 &   23.7 &   23.1  &   51.9  &   49.3  &   45.1  &   51.0   \\
I & J033157.1$-$274525.0 &  52.9881  &  $-$27.7570 &   26.8 &   26.6 &   24.7 &   23.5  &   95.2  &   95.5  &   95.0  &   94.6   \\
J & J033203.1$-$274543.3 &  53.0132  &  $-$27.7620 &   27.4 &   25.5 &   24.5 &   23.6  &   98.3  &   97.2  &   99.0  &   98.4   \\
K & J033206.9$-$274720.6 &  53.0290  &  $-$27.7891 &  \nodata &   27.8 &   25.0 &   23.5  &   98.3  &   97.9  &   99.0  &   98.4   \\
L & J033202.2$-$274859.7 &  53.0093  &  $-$27.8166 &   26.4 &   25.5 &   24.7 &   24.1  &   98.3  &   97.2  &   99.0  &   98.4   \\
M & J033204.1$-$274930.2 &  53.0175  &  $-$27.8251 &   27.0 &   26.2 &   25.2 &   24.4  &   98.3  &   97.9  &   99.0  &   98.4   \\
N & J033159.5$-$275020.2 &  52.9981  &  $-$27.8390 &   26.6 &   27.0 &   24.8 &   23.4  &   98.2  &   97.2  &   98.8  &   98.4   \\
O & J033152.7$-$274928.6 &  52.9699  &  $-$27.8246 &  \nodata &   26.3 &   24.8 &   23.9  &   93.1  &   96.0  &   95.1  &   97.0   \\
P & J033149.0$-$274950.3 &  52.9543  &  $-$27.8307 &   27.1 &   27.2 &   25.4 &   24.4  &   92.0  &   89.0  &   83.1  &   83.6   \\
Q & J033142.8$-$274938.0 &  52.9287  &  $-$27.8272 &   27.3 &   27.0 &   25.5 &   24.3  &   82.0  &   77.8  &   74.0  &   66.8   \\
R & J033144.4$-$275021.1 &  52.9353  &  $-$27.8392 &   30.3 &   29.0 &   25.6 &   24.4  &   87.0  &   83.7  &   78.9  &   69.9   \\
S & J033147.3$-$275218.6 &  52.9474  &  $-$27.8719 &   \nodata &   28.0 &   26.0 &   24.4  &   90.5  &   90.1  &   85.5  &   83.8   \\
T & J033202.1$-$275242.1 &  53.0088  &  $-$27.8784 &   26.8 &   26.6 &   25.1 &   24.2  &   98.0  &   97.9  &   99.0  &   98.4   \\
U & J033214.9$-$275302.0 &  53.0625  &  $-$27.8839 &   25.8 &   25.2 &   24.1 &   23.3  &  100.0  &  100.0  &  100.0  &  100.0   \\
V & J033217.2$-$275353.4 &  53.0720  &  $-$27.8982 &   26.5 &   25.9 &   25.0 &   24.4  &  100.0  &  100.0  &  100.0  &  100.0   \\
W & J033202.4$-$275333.4 &  53.0102  &  $-$27.8926 &   25.7 &   26.0 &   24.8 &   23.9  &   98.0  &   97.9  &   99.0  &   98.7   \\
X & J033200.4$-$275459.2 &  53.0018  &  $-$27.9165 &   28.3 &   27.3 &   25.4 &   23.9  &   97.1  &   97.2  &   97.6  &   98.4   \\
Y & J033151.5$-$275453.6 &  52.9650  &  $-$27.9149 &   25.8 &   25.3 &   24.3 &   23.9  &   71.5  &   78.3  &   74.5  &   82.8   \\
Z & J033149.7$-$275440.9 &  52.9574  &  $-$27.9114 &   26.1 &   25.6 &   24.5 &   24.1  &   54.4  &   58.0  &   62.1  &   65.5   \\
AA & J033150.1$-$275506.2 &  52.9588  &  $-$27.9184 &   29.2 &   28.6 &   25.5 &   24.3  &   45.4  &   51.3  &   58.8  &   58.9   \\
BB & J033158.6$-$275732.2 &  52.9946  &  $-$27.9590 &   \nodata &   28.1 &   25.3 &   24.4  &   53.5  &   54.3  &   64.9  &   72.8   \\
CC & J033211.6$-$275735.7 &  53.0486  &  $-$27.9599 &   26.8 &   25.3 &   24.4 &   23.5  &   92.1  &   91.6  &   91.9  &   93.6   \\
DD & J033216.0$-$275703.0 &  53.0669  &  $-$27.9509 &   \nodata &   26.2 &   25.3 &   24.4  &   92.8  &   94.4  &   93.2  &   95.9   \\
EE & J033225.5$-$275706.2 &  53.1063  &  $-$27.9517 &   26.5 &   26.8 &   25.2 &   24.4  &   91.1  &   92.3  &   94.2  &   94.5   \\
FF & J033242.9$-$275511.7 &  53.1789  &  $-$27.9199 &   28.8 &   26.1 &   25.0 &   24.2  &   95.8  &   94.7  &   98.3  &   97.9   \\
GG & J033246.8$-$275448.8 &  53.1954  &  $-$27.9136 &   26.9 &   27.7 &   25.3 &   24.2  &   96.5  &   94.2  &   97.9  &   96.6   \\
HH & J033235.5$-$275447.0 &  53.1481  &  $-$27.9131 &   26.1 &   25.3 &   24.5 &   23.8  &  100.0  &  100.0  &  100.0  &  100.0   \\
II & J033232.0$-$275326.6 &  53.1337  &  $-$27.8907 &   26.4 &   26.2 &   25.1 &   24.4  &  100.0  &  100.0  &  100.0  &  100.0   \\
JJ & J033231.4$-$275137.6 &  53.1310  &  $-$27.8605 &   27.0 &   26.3 &   25.1 &   24.1  &  100.0  &  100.0  &  100.0  &  100.0   \\
KK & J033223.9$-$275031.8 &  53.0997  &  $-$27.8422 &   25.8 &   26.4 &   24.8 &   23.8  &  100.0  &  100.0  &  100.0  &  100.0   \\
LL & J033224.4$-$275034.5 &  53.1019  &  $-$27.8429 &   30.1 &   27.4 &   25.4 &   24.0  &  100.0  &  100.0  &  100.0  &  100.0   \\
MM & J033237.2$-$275013.2 &  53.1554  &  $-$27.8370 &   25.2 &   24.5 &   23.7 &   23.1  &  100.0  &  100.0  &  100.0  &  100.0   \\
NN & J033244.9$-$275005.0 &  53.1871  &  $-$27.8347 &   27.5 &   26.0 &   24.8 &   24.0  &  100.0  &  100.0  &  100.0  &  100.0   \\
OO & J033221.1$-$274950.6 &  53.0881  &  $-$27.8307 &   25.7 &   26.0 &   25.2 &   24.4  &  100.0  &  100.0  &  100.0  &  100.0   \\
PP & J033228.9$-$274908.4 &  53.1208  &  $-$27.8190 &   25.6 &   24.8 &   23.9 &   23.0  &  100.0  &  100.0  &  100.0  &  100.0   \\
QQ & J033235.3$-$274920.2 &  53.1472  &  $-$27.8223 &   26.3 &   26.4 &   25.3 &   24.5  &  100.0  &  100.0  &  100.0  &  100.0   \\
RR & J033235.9$-$274850.3 &  53.1499  &  $-$27.8140 &   25.0 &   23.9 &   23.1 &   22.5  &  100.0  &  100.0  &  100.0  &  100.0   \\
SS & J033220.8$-$274822.7 &  53.0869  &  $-$27.8063 &   27.9 &   27.3 &   24.9 &   23.8  &  100.0  &  100.0  &  100.0  &  100.0   \\
TT & J033234.6$-$274727.3 &  53.1445  &  $-$27.7909 &   28.8 &   26.7 &   25.4 &   24.4  &  100.0  &  100.0  &  100.0  &  100.0   \\
UU & J033238.4$-$274725.1 &  53.1603  &  $-$27.7903 &   26.7 &   26.8 &   25.1 &   24.3  &  100.0  &  100.0  &  100.0  &  100.0   \\
VV & J033237.7$-$274641.5 &  53.1574  &  $-$27.7782 &   26.7 &   26.2 &   25.1 &   24.2  &  100.0  &  100.0  &  100.0  &  100.0   \\
WW & J033246.1$-$274529.6 &  53.1924  &  $-$27.7582 &   27.3 &   26.3 &   25.2 &   24.4  &   95.7  &   95.2  &   97.2  &   91.4   \\
XX & J033253.2$-$274644.3 &  53.2218  &  $-$27.7790 &   26.4 &   29.2 &   25.6 &   23.9  &   87.8  &   74.6  &   76.3  &   74.7   \\
\cutinhead{Rejected Candidates}
 & J033145.7$-$275003.5 &  52.9406  &  $-$27.8343 &   26.7 &   26.9 &   25.5 &   24.3  &   89.3  &   86.4  &   81.3  &   71.2   \\
 & J033203.4$-$275059.5 &  53.0143  &  $-$27.8499 &   26.4 &   27.5 &   25.2 &   23.9  &   98.3  &   97.9  &   99.0  &   98.4   \\
& J033206.8$-$274208.6 &  53.0286  &  $-$27.7024 &   25.6 &   25.9 &   24.9 &   24.3  &   54.1  &   60.5  &   49.2  &   53.8   \\
 & J033207.4$-$274400.4 &  53.0310  &  $-$27.7335 &   25.8 &   25.3 &   23.8 &   22.8  &   98.3  &   97.9  &   99.0  &   98.3   \\
& J033215.2$-$275039.8 &  53.0636  &  $-$27.8444 &   26.0 &   26.0 &   24.9 &   24.0  &  100.0  &  100.0  &  100.0  &  100.0   \\
 & J033215.3$-$275044.1 &  53.0639  &  $-$27.8456 &   25.2 &   25.6 &   23.9 &   22.8  &  100.0  &  100.0  &  100.0  &  100.0   \\
 & J033254.6$-$275008.3 &  53.2278  &  $-$27.8356 &   26.9 &   27.5 &   25.1 &   23.6  &   98.6  &   96.7  &  100.0  &   94.5   \\
 & J033258.4$-$274955.4 &  53.2436  &  $-$27.8321 &   26.1 &   26.1 &   24.2 &   22.6  &   91.6  &   86.1  &   95.4  &   82.7   \\
\enddata 
\tablenotetext{*}{T are ratios of the observed exposure time at the location of the source to maximum exposure time for the field. See \cite{Hoversten10} for maximum exposure times.}
 \label{tab:raw}
\end{deluxetable*} 
\begin{deluxetable*}{clllccccccc}
\tabletypesize{\scriptsize}
\setlength{\tabcolsep}{0.07in}
\tablecolumns{11} 
\tablewidth{0pt} 
\tablecaption{Derived Quantities} 
\tablehead{\colhead{ID} & \colhead{J2000 ID} & \colhead{RA} & \colhead{Dec} & \colhead{$z$} & \colhead{$z$ ref\tablenotemark{a}} & \colhead{SFR$_{\rm UV}$} & \colhead{SFR$_{\rm tot}$\tablenotemark{b}} & \colhead{Log (M$_*$)} & \colhead{A$_{\rm FUV}$} & \colhead{Phot\tablenotemark{c}} \\
& & & & & & \colhead{(M$_\odot$ yr$^{-1}$)} & \colhead{(M$_\odot$ yr$^{-1}$)} &  \colhead{(M$_\odot$)}& &
}
\startdata
\cutinhead{LBG Sample}
A & J033226.8$-$274156.6 &  53.1117  &  $-$27.6991 &   1.614 $^{\rm s}$ &  1  &      5.3  &   26.2  &    9.4  &    1.7 &      2,4   \\
B & J033230.3$-$274241.0 &  53.1264  &  $-$27.7114 &   1.890 $^{\rm s}$ &  1  &      3.8  &   27.6  &    9.8  &    2.0 &    2,3,4   \\
C & J033228.0$-$274249.9 &  53.1167  &  $-$27.7139 &    1.9 $^{\rm p}$ &  2  &      2.5  &    3.2  &    9.8  &    0.2 &      2,4   \\
D & J033230.6$-$274345.6 &  53.1279  &  $-$27.7293 &    1.3 $^{\rm p}$ &  4  &      1.4  &    5.0  &    9.2  &    1.4 &      2,4   \\
E & J033226.8$-$274425.0 &  53.1119  &  $-$27.7403 &   1.613 $^{\rm s}$ &  1  &      2.6  &   53.4  &    9.1  &    3.0 &      2,4   \\
F & J033218.8$-$274500.0 &  53.0784  &  $-$27.7500 &   1.572 $^{\rm s}$ &  1  &      1.5  &    6.8  &    9.3  &    1.6 &      2,4   \\
G & J033219.8$-$274516.2 &  53.0827  &  $-$27.7545 &    1.3 $^{\rm p}$ &  4  &      0.9  &   14.0  &    9.1  &    2.7 &      2,4   \\
H & J033149.4$-$274456.3 &  52.9560  &  $-$27.7490 &    1.2 $^{\rm p}$ &  4  &      3.5   &   (3.5)  &     9.7  &   \nodata &        3   \\
I & J033157.1$-$274525.0 &  52.9881  &  $-$27.7570 &   1.237 $^{\rm s}$ &  1  &      2.3   &   (2.3)  &     9.5  &   \nodata &        3   \\
J & J033203.1$-$274543.3 &  53.0132  &  $-$27.7620 &   0.707 $^{\rm s}$ &  1  &      0.5  &   11.3  &    9.6  &    3.1 &        3   \\
K & J033206.9$-$274720.6 &  53.0290  &  $-$27.7891 &    1.9 $^{\rm p}$ &  2  &    \nodata     &     \nodata   &    \nodata   &   \nodata &            \\
L & J033202.2$-$274859.7 &  53.0093  &  $-$27.8166 &    1.3 $^{\rm p}$ &  4  &      1.4   &   (1.4)  &     9.6  &   \nodata &        4   \\
M & J033204.1$-$274930.2 &  53.0175  &  $-$27.8251 &    1.2 $^{\rm p}$ &  4  &      1.0  &    9.4  &    8.9  &    2.3 &        4   \\
N & J033159.5$-$275020.2 &  52.9981  &  $-$27.8390 &    1.1 $^{\rm p}$ &  4  &      1.2   &   (1.2)  &     9.7  &   \nodata &        3   \\
O & J033152.7$-$274928.6 &  52.9699  &  $-$27.8246 &    1.3 $^{\rm p}$ &  3  &    \nodata     &     \nodata   &    \nodata   &   \nodata &            \\
P & J033149.0$-$274950.3 &  52.9543  &  $-$27.8307 &    0.7 $^{\rm p}$ &  4  &      0.2   &   (0.2)  &     8.1  &   \nodata &        4   \\
Q & J033142.8$-$274938.0 &  52.9287  &  $-$27.8272 &    1.2 $^{\rm p}$ &  4  &      1.1   &   (1.1)  &     9.8  &   \nodata &        3   \\
R & J033144.4$-$275021.1 &  52.9353  &  $-$27.8392 &    1.0 $^{\rm p}$ &  4  &      0.6   &   (0.6)  &     8.7  &   \nodata &        4   \\
S & J033147.3$-$275218.6 &  52.9474  &  $-$27.8719 &    1.2 $^{\rm p}$ &  4  &    \nodata     &     \nodata   &    \nodata   &   \nodata &            \\
T & J033202.1$-$275242.1 &  53.0088  &  $-$27.8784 &    0.7 $^{\rm p}$ &  4  &      0.4   &   (0.4)  &     8.5  &   \nodata &        4   \\
U & J033214.9$-$275302.0 &  53.0625  &  $-$27.8839 &   1.359 $^{\rm s}$ &  1  &      3.1  &    6.5  &    9.3  &    0.8 &      2,4   \\
V & J033217.2$-$275353.4 &  53.0720  &  $-$27.8982 &   1.350 $^{\rm s}$ &  1  &      1.1  &    5.7  &    9.4  &    1.7 &    2,3,4   \\
W & J033202.4$-$275333.4 &  53.0102  &  $-$27.8926 &    0.8 $^{\rm p}$ &  4  &      0.6   &   (0.6)  &     8.5  &   \nodata &        4   \\
X & J033200.4$-$275459.2 &  53.0018  &  $-$27.9165 &    1.2 $^{\rm p}$ &  4  &      0.9   &   (0.9)  &     9.0  &   \nodata &        4   \\
Y & J033151.5$-$275453.6 &  52.9650  &  $-$27.9149 &    1.7 $^{\rm p}$ &  4  &      3.0   &   (3.0)  &    10.7  &   \nodata &        4   \\
Z & J033149.7$-$275440.9 &  52.9574  &  $-$27.9114 &    1.2 $^{\rm p}$ &  4  &      2.0   &   (2.0)  &     9.4  &   \nodata &        3   \\
AA & J033150.1$-$275506.2 &  52.9588  &  $-$27.9184 &    2.0 $^{\rm p}$ &  4  &      2.8   &   (2.8)  &     9.6  &   \nodata &        4   \\
BB & J033158.6$-$275732.2 &  52.9946  &  $-$27.9590 &    1.6 $^{\rm p}$ &  4  &    \nodata     &     \nodata   &    \nodata   &   \nodata &            \\
CC & J033211.6$-$275735.7 &  53.0486  &  $-$27.9599 &    1.6 $^{\rm p}$ &  4  &      2.9  &   26.4  &   11.1  &    2.2 &        4   \\
DD & J033216.0$-$275703.0 &  53.0669  &  $-$27.9509 &    0.5 $^{\rm p}$ &  4  &    \nodata     &     \nodata   &    \nodata   &   \nodata &            \\
EE & J033225.5$-$275706.2 &  53.1063  &  $-$27.9517 &    1.3 $^{\rm p}$ &  4  &      1.1  &   16.7  &    9.0  &    2.7 &        4   \\
FF & J033242.9$-$275511.7 &  53.1789  &  $-$27.9199 &    1.4 $^{\rm p}$ &  4  &      1.6  &  212.8  &    9.2  &    4.9 &      2,4   \\
GG & J033246.8$-$275448.8 &  53.1954  &  $-$27.9136 &   1.552 $^{\rm s}$ &  1  &      2.1  &    7.5  &    9.3  &    1.3 &    2,3,4   \\
HH & J033235.5$-$275447.0 &  53.1481  &  $-$27.9131 &    1.2 $^{\rm p}$ &  4  &      1.9  &    2.9  &    9.4  &    0.5 &    2,3,4   \\
II & J033232.0$-$275326.6 &  53.1337  &  $-$27.8907 &   0.998 $^{\rm s}$ &  1  &      0.3  &    2.0  &    8.5  &    2.0 &      2,4   \\
JJ & J033231.4$-$275137.6 &  53.1310  &  $-$27.8605 &   1.382 $^{\rm s}$ &  1  &      1.4  &    6.2  &    9.7  &    1.5 &      2,4   \\
KK & J033223.9$-$275031.8 &  53.0997  &  $-$27.8422 &    1.0 $^{\rm p}$ &  4  &      0.5  &   25.4  &    8.9  &    3.8 &      2,4   \\
LL & J033224.4$-$275034.5 &  53.1019  &  $-$27.8429 &   1.552$^{\rm s}$ &  1  &      1.3  &    8.6  &    9.1  &    1.9 &    2,3,4   \\
MM & J033237.2$-$275013.2 &  53.1554  &  $-$27.8370 &   1.389 $^{\rm s}$ &  1  &      3.1  &   41.1  &    9.3  &    2.6 &    2,3,4   \\
NN & J033244.9$-$275005.0 &  53.1871  &  $-$27.8347 &   1.296 $^{\rm s}$ &  1  &      1.1  &    5.1  &    9.0  &    1.6 &      2,4   \\
OO & J033221.1$-$274950.6 &  53.0881  &  $-$27.8307 &   0.965 $^{\rm s}$ &  1  &      0.5  &    6.3  &    9.7  &    2.6 &    2,3,4   \\
PP & J033228.9$-$274908.4 &  53.1208  &  $-$27.8190 &   1.094 $^{\rm s}$ &  1  &      1.6  &   68.5  &   10.6  &    3.6 &    2,3,4   \\
QQ & J033235.3$-$274920.2 &  53.1472  &  $-$27.8223 &   0.666 $^{\rm s}$ &  1  &      0.3  &    2.6  &    8.8  &    2.1 &    2,3,4   \\
RR & J033235.9$-$274850.3 &  53.1499  &  $-$27.8140 &   1.309 $^{\rm s}$ &  1  &      5.8  &   48.6  &   10.1  &    2.1 &    2,3,4   \\
SS & J033220.8$-$274822.7 &  53.0869  &  $-$27.8063 &    1.9 $^{\rm p}$ &  2  &      3.1  &    6.7  &    9.1  &    0.9 &      2,4   \\
TT & J033234.6$-$274727.3 &  53.1445  &  $-$27.7909 &   1.438 $^{\rm s}$ &  1  &      1.3  &    7.9  &    8.8  &    1.9 &      2,4   \\
UU & J033238.4$-$274725.1 &  53.1603  &  $-$27.7903 &    1.9 $^{\rm p}$ &  2  &      2.2  &    9.8  &    9.5  &    1.6 &      2,4   \\
VV & J033237.7$-$274641.5 &  53.1574  &  $-$27.7782 &   1.307 $^{\rm s}$ &  1  &      0.8  &    1.4  &    8.7  &    0.6 &      2,4   \\
WW & J033246.1$-$274529.6 &  53.1924  &  $-$27.7582 &   1.298 $^{\rm s}$ &  1  &      0.6  &    4.0  &    9.2  &    1.9 &      2,4   \\
XX & J033253.2$-$274644.3 &  53.2218  &  $-$27.7790 &    1.9 $^{\rm p}$ &  2  &      2.4  &    3.5  &   11.0  &    0.4 &        4   \\
\cutinhead{Rejected Candidates}
 & J033145.7$-$275003.5 &  52.9406  &  $-$27.8343 &    0.0 $^{\rm p}$ &  4  &      0.0   &   (0.0)  &     6.3  &   \nodata &        4   \\
 & J033203.4$-$275059.5 &  53.0143  &  $-$27.8499 &   0.00 $^{\rm s}$ &  1  &    \nodata     &     \nodata   &    \nodata   &   \nodata &        4   \\
 & J033206.8$-$274208.6 &  53.0286  &  $-$27.7024 &   0.286 $^{\rm s}$ &  1  &      0.0  &    0.4  &    9.3  &    3.2 &    2,3,4   \\
& J033207.4$-$274400.4 &  53.0310  &  $-$27.7335 &  $-$99.0 $^{\rm p}$ &  4  &    \nodata     &     \nodata   &    \nodata   &   \nodata &            \\
 & J033215.2$-$275039.8 &  53.0636  &  $-$27.8444 &   0.246 $^{\rm s}$ &  1  &      0.0  &    0.9  &    8.8  &    3.9 &    2,3,4   \\
 & J033215.3$-$275044.1 &  53.0639  &  $-$27.8456 &   0.230 $^{\rm s}$ &  1  &      0.0  &    0.4  &   10.0  &    2.8 &    2,3,4   \\
 & J033254.6$-$275008.3 &  53.2278  &  $-$27.8356 &    0.0 $^{\rm p}$ &  4  &    \nodata     &     \nodata   &    \nodata   &   \nodata &        3   \\
& J033258.4$-$274955.4 &  53.2436  &  $-$27.8321 &   0.00 $^{\rm s}$ &  1  &    \nodata     &     \nodata   &    \nodata   &   \nodata &        3   \\
\enddata
\tablenotetext{a}{References: (1) Spectroscopic redshift from the literature, (2) ACS-GOODS photometric redshift \citep{DahlenTomas2007}), (3) photometric redshift from MUSYC K-selected sample \citep{musycK}, (4) photometric redshift from COMBO-17 \citep{combo17}}.
\tablenotetext{b}{SFR$_{\rm tot}$=SFR$_{\rm TIR}$+SFR$_{\rm UV}$ -- Parentheses mark where no IR data is available, and SFR$_{\rm tot}$=SFR$_{\rm UV}$. }.
\tablenotetext{c}{Photometric catalogs used in the SED fit-- references are same as for the redshift determination (see above, note a)}.
\tablenotetext{s}{Spectroscopic redshift  -- quoted to the typical precision for the spectroscopic redshifts reported in the GOODS spectroscopic catalog: {\texttt http://www.eso.org/sci/activities/garching/projects/goods/MASTERCAT\_v2.0.dat}
}.
\tablenotetext{p}{Photometric redshift-- $\langle \Delta z/(1+z)\rangle \sim 0.1$ for these galaxies.
}.
\label{tab:derived}
\end{deluxetable*} 

%
   
To summarize, we reject J033145.7$-$275003.5,  J033206.8$-$274208.6, J033207.4$-$274400.4, J033215.3$-$275044.1, J033254.6$-$275008.3 and J033258.4$-$274955.4 because we believe these are unlikely to be 0.5$<z<2.0$ LBGs; we reject J033203.4$-$275059.5 and J033215.2$-$275039.8 since they are difficult to study because of insufficient spatial resolution of UVOT to avoid source confusion from bright objects nearby (star in former case and elliptical galaxy in latter) and missing optical counterpart in latter case. Therefore, of the 58 candidates:  6 appear to be low redshift ($z<0.5$) interlopers ($\sim$10\%) and 2 are undetermined sources. 

\subsection{AGN Contribution}
We match our 58 candidates with sources in the Chandra 2 Ms CDF-S catalog \citep{Luo} in order to determine AGN contribution in these sources. We find 3 matches (within 2\arcsec): LBGs N, RR and UU. LBG N was detected in both hard (2-8 keV) and full (0.5-8 keV) X-ray bands, with luminosities $\sim10^{43} {\rm~erg~s^{-1}}$. LBG RR was detected in all bands: hard (2$-$8 keV), soft (0.5$-$2 keV), and full (0.5$-$8 keV) X-ray bands, with luminosities 2.5$\times10^{42} {\rm~erg~s^{-1}}$, 1.5$\times10^{42} {\rm~erg~s^{-1}}$ and 4$\times10^{42} {\rm~erg~s^{-1}}$, respectively. LBG UU was only detected in the hard X-ray band (2-8 keV) with luminosity of 4$\times10^{42}{\rm~erg~s^{-1}}$. Based on the study by \cite{Silverman}, low-luminosity AGNs ($L_{X}<10^{44}{\rm~erg~s^{-1}}$) do not affect the optical magnitudes or colors of the host galaxy; therefore, we do not expect that the presence of low-luminosity AGN in these three LBGs will have significant effect on their analysis. 

The AGN fraction for this sample of 0.5$<z<$2.0 LBGs is $\sim5-6$\%, consistent with AGN fractions of other star-forming galaxies in CDFS \citep[$z<$1 late-type galaxies and $z=$3 LBGs;][]{Lehmer08}.

\subsection{Morphology Analysis}
To measure the morphology of the galaxies, we use the \HST/ACS V$_{606}$-band images from either the Galaxy Evolution through Morphology and SEDs \citep[GEMS;][]{gems} or GOODS \citep[][]{giaGOODS} South datasets. These images are shown in Figure \ref{fig:mapsbw} and \ref{fig:mapsrejbw}. However we apply the morphology analysis only to the 50 LBGs (not to the 8 rejected candidates). 

We have used the methodology described by \cite{Zamojski} to derive morphological parameters, including concentration (C), asymmetry (A), clumpiness (S), Gini (G) and M$_{20}$. In our analysis, we use only C, G, and $M_{20}$ to compare with other UV-selected galaxies. We will describe those parameters briefly here and refer readers to other relevant papers \citep{Abraham96, Abraham2003, Conselice2000, Conselice03, Lotz04} for further details.

C measures the concentration of the light distribution, C$\equiv \log \frac{\rm r_{80\%}}{\rm r_{20\%}}$, where r$_{80\%}$ and r$_{20\%}$ are the radii containing 80\% of the light and 20\% of the light, respectively. The Gini parameter, G, measures the inequality of the light distribution. This is similar to C, but doesn't depend on the location of the source centroid. G ranges from 0, where all the pixels have uniform intensity, to 1, where most of the flux is concentrated in a single pixel. $M_{20}$ is the normalized second order moment of the brightest 20\% of the galaxy's flux: $M_{20}=\log \frac{\Sigma_if_ir_i^2}{M_{tot}}$, where $f_i$ and $r_i$ are the flux and distance from center of the $i$th pixel, summed over the pixels in order of decreasing brightness until 20\% of the total flux is reached and $M_{tot}$ is the total flux over all the pixels. Together, these parameters effectively identify mergers and bulge-like morphologies. 
   
   \subsection{Derived Dust Attenuation and Total SFR Using MIPS Data}
40 galaxies (34 LBGs and 6 rejected candidates) also appeared in the extended CDF-S Spitzer Multiband Imaging Photometer \citep[MIPS;][]{MIPS} infrared catalog, providing 24$\mu$m photometry \citep{Chary2004, Charyproc}. Using the code from \citet{CharyElbaz}, we estimate the total infrared luminosity and IR-based SFRs (except that we modify the assumed Salpeter IMF to Kroupa IMF), SFR$_{TIR}$, for these galaxies. \cite{Elbaz2010} use this code to compare L$_{\rm TIR}$ estimated from 24$\mu$m alone (L$_{\rm TIR, 24\mu m}$) to the \textit{Herschel}-derived L$_{\rm TIR, {\it Herschel}}$. They find that there is relatively good agreement for Log${\rm (L_{TIR, 24\mu m}/L_\odot}) < 12 $ for $0.5<z<1.5$ galaxies, while L$_{\rm TIR}$ (and therefore SFR$_{\rm TIR}$) is overestimated when calculated from the 24$\mu$m flux alone for more luminous 24$\mu$m sources. Only one of our sources, `FF',  is more luminous at Log$({\rm L_{TIR, 24\mu m}/L_\odot}) =12.6$. According to \cite{Elbaz2010}, the $Herschel$ value should be Log${\rm(L_{\rm TIR, {\it Herschel}}/L_\odot})\approx12.3$, used for values in Table \ref{tab:derived}.

We combine these infrared SFRs with dust-uncorrected UV-derived SFRs, SFR$_{\rm UV}$, to calculate total SFRs (SFR$_{\rm tot}$=SFR$_{\rm TIR}$+SFR$_{\rm UV}$), modifying the \cite{Kennicutt} UV SFR relation from Salpeter IMF into Kroupa IMF. Furthermore, we calculate FUV attenuation, A$_{\rm FUV}$, following \cite{Burgarella2005a}:
\begin{align}
\rm A_{FUV}  =  &-0.028[\rm Log(F_{IR}/F_{UV})]^3 \notag \\ 
& + 0.392[\rm Log(F_{IR}/F_{UV})]^2 \notag \\
& + 1.494[\rm Log(F_{IR}/F_{UV})] + 0.546
\end{align}
where the F$_{\rm IR}$ and F$_{\rm UV}$ are the infrared and ultraviolet fluxes. Table \ref{tab:derived} presents the derived SFR$_{\rm tot}$ and A$_{\rm FUV}$ parameters for the candidates. We note that the dust law (\ie the dependence of dust attenuation curve on $\lambda$, k($\lambda$)) in \cite{Burgarella2005a} differs from the one we have been assuming in earlier sections \citep{Calzetti94}. In their work, \cite{Burgarella2005a} fit a range of slopes, $\alpha$, and 2175 \AA~dust bump strengths, A$_{\rm bump}$ to the dust law given by: $\rm k(\lambda)=\lambda^{\alpha}+ A_{bump} exp^{[(\lambda-2175\AA)/\sigma^2]}$; they find that galaxies span a range of slopes and dust bump strengths and the best estimate for A$_{\rm FUV}$ comes from the far-infrared (FIR). Therefore, when FIR data is unavailable (\eg in Figure \ref{fig:sel}) we assume the simple dust law from \cite{Calzetti94}. 

\subsection{Comparison Samples}\label{sec:comp}
We select two comparison samples from the ACS-GOODS catalog. The first comparison sample is comprised of all the $0.5<z<2$ galaxies, according to the photometric redshifts (with uncertainties $\langle\Delta z/(1+z)\rangle\sim 0.06$) from the ACS-GOODS catalog \citep[see][]{Dahlen2010}, and we refer to these 8247 galaxies as the ``$z\sim1$ ACS'' sample. 

The second comparison sample is derived from this  $z\sim1$ ACS parent sample with the intention of simulating a larger and fainter LBG sample.  Similar to our analysis of the UVOT-selected LBGs, we use {\it kcorrect} to SED-fit the ACS-GOODs photometry. We project the best-fit SED (in the observed frame) onto the UVOT filters to simulate UVOT observations for these galaxies. Then we select LBGs, using Equations \ref{eqn:lbg2} $-$ \ref{eqn:lbg4} and we replace Equation \ref{eqn:lbg1} with $20.75<u<26.5$, extending to fainter galaxies. We refer to this sample of 1630 galaxies  as the ``simulated LBG'' sample. 

\begin{figure*}
\includegraphics[width=7.0in]{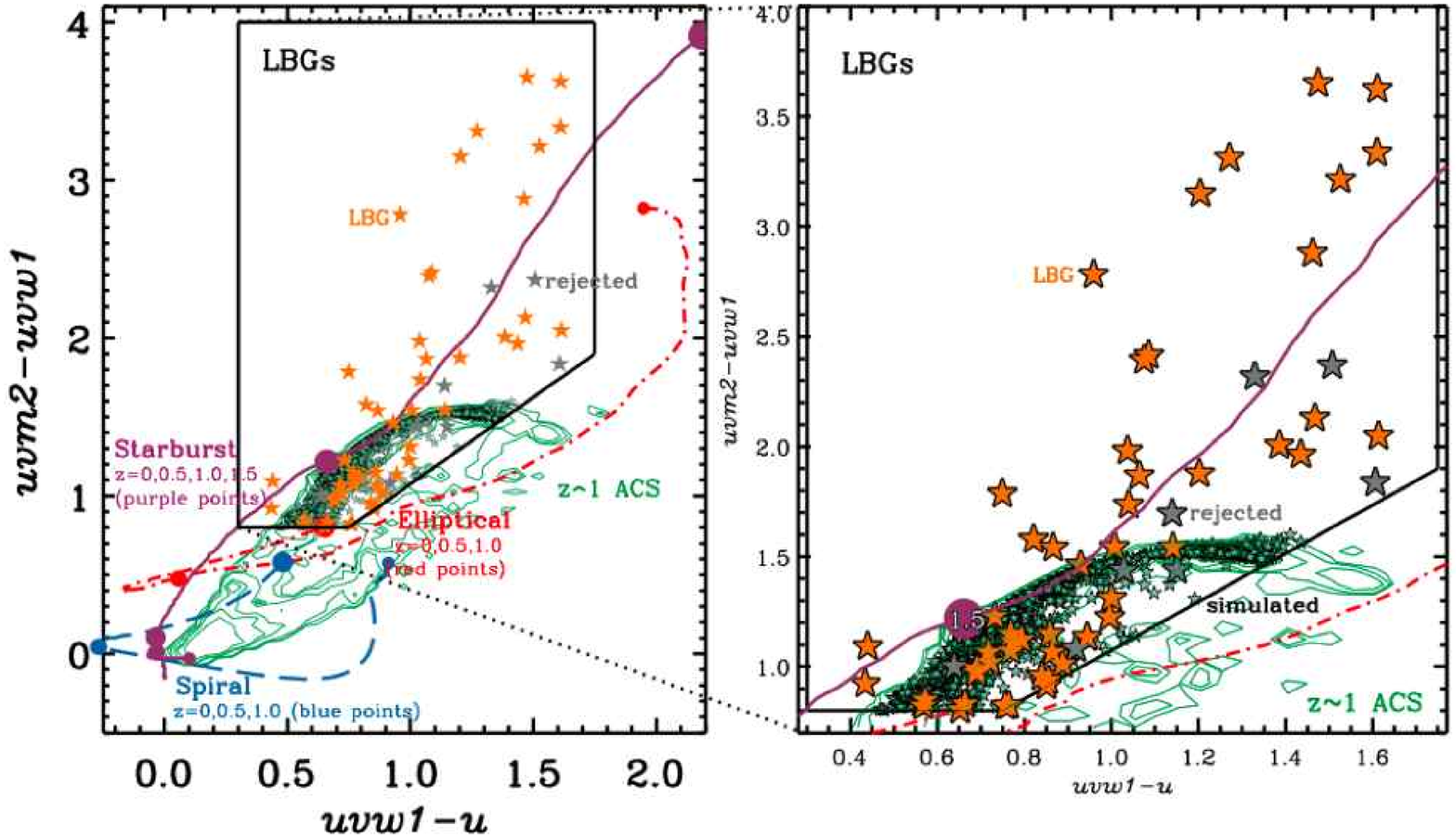}
\caption{Same as Figure \ref{fig:sel} (with omission of the gray CDF-S UVOT catalog points, dark gray outline of region populated by stars, light purple region outlining range of dust attenuation values for starburst track, and LBG candidate names for clarity), but showing the comparison samples: dark green contours showing $z\sim1$ ACS sample and light green stars, outlined in black, marking the simulated LBG sample. }\label{fig:lbgsel_full}
\vspace{0.08in}
\end{figure*}

Around 7\% of simulated LBGs have best-fit spectra dominated by older stellar populations, resembling less star-forming galaxies. In Figure \ref{fig:lbgsel_full}, we show the simulated UVOT colors for the $z\sim1$ ACS sample and the simulated LBG sample in open green circles and light green stars, respectively.  

\section{Results}\label{sec:results}  
Recent technological advances in ultraviolet detectors have provided us with the opportunity to locate and study $z\sim1$ LBGs and to compare these LBGs with other galaxy populations at these redshifts and with the higher redshift ($z>3$) LBG population. In this study, we test the utility of selecting LBGs using \Swift~UVOT. \cite{Hathi2010} and \cite{Oesch2010} have used \HST~WFC3 to select hundreds of LBGs in this regime, benefitting from the sensitivity of WFC3 to select galaxies $\approx 2$ magnitudes deeper than our sample. 
While the \Swift~UVOT is not as sensitive as WFC3, the larger FOV of \Swift~and the number of other deeply observed GRB fields (there are currently 40 fields with $>$200ks of exposure in the UV filters) offer advantages for studying the bright subset of LBGs with \Swift~UVOT. 

In the following subsections, we discuss the properties the UVOT-selected LBGs and compare this sample with the comparison samples and with other LBGs at $z\sim$3 and $z\sim$0.2 LBAs.

\subsection{Morphology Results}
\begin{figure*}
\begin{center}
  \includegraphics[width=6.5in]{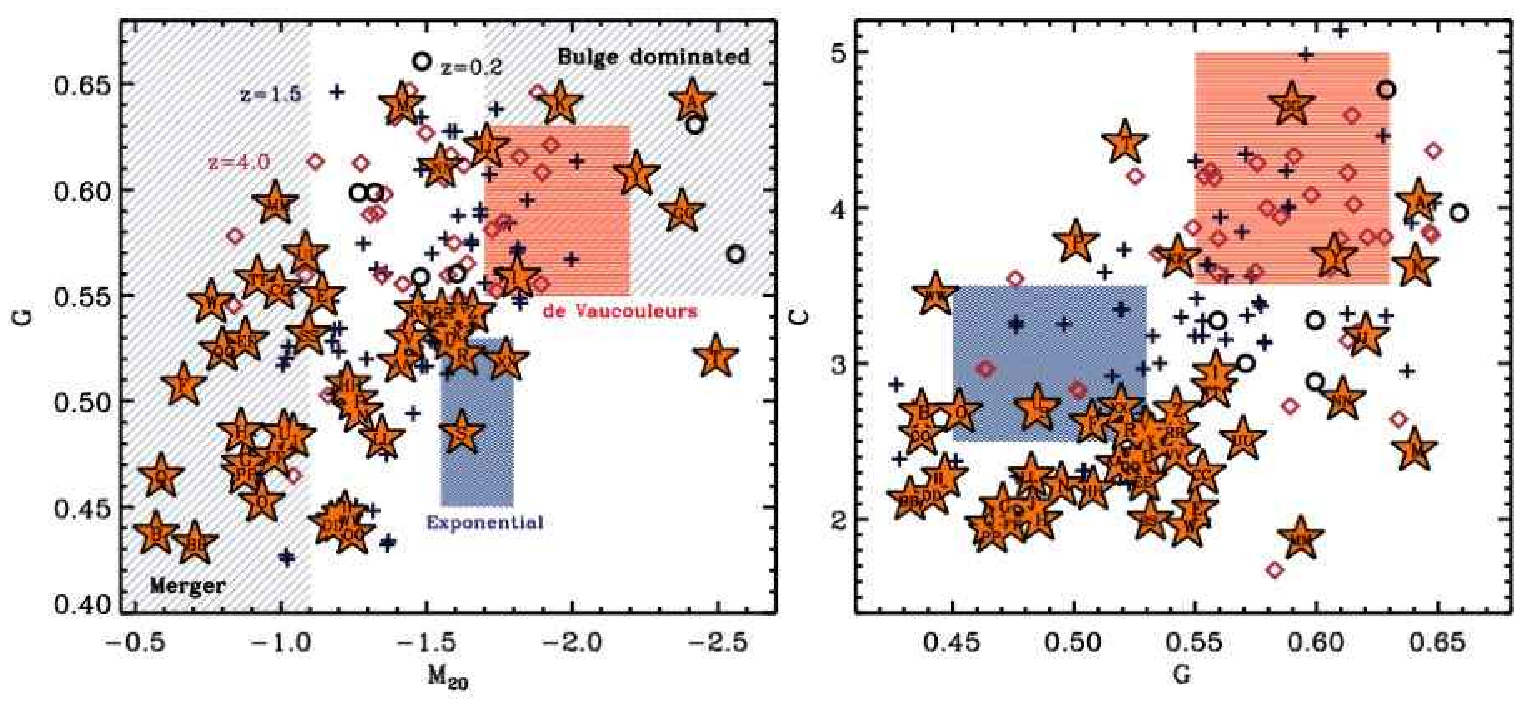}
    \end{center}
  \caption{LBGs selected in this study (orange stars) appear similar in morphology (Gini, M$_{20}$, and concentration) to other LBG or star-forming populations at different redshifts: emission-line selected galaxies at $z=1.5$ (blue crosses) from \cite{Lotz06}, LBGs at $z=4.0$ (magenta diamonds) from \cite{Lotz06}, and low redshift $z=0.2$ LBAs (black open circles) from \cite{rod}. As shown in the left panel, a significant fraction of these LBGs ($\geq35$\%) resembles mergers (the hatched region on left).  In this panel we also mark the region inhabited by bulge-dominated galaxies (upper right hatched region); in both panels, the red and navy shaded regions mark the simulated parameters for galaxies following de Vaucouleurs and exponential light profiles. Morphology is determined from rest-frame FUV images for the \cite{Lotz06} and \cite{rod} samples, and from rest-frame NUV image for LBGs (or observed V-band, since B-band was unavailable for many of the galaxies). }
   \label{fig:morph}
\vspace{0.05in}
   \end{figure*}

Based on the optical images (Figure \ref{fig:mapsbw}) most of the LBGs appear compact, blue and clumpy with mean half light radii of 2.3 kpc and high surface brightnesses. However, LBGs B, C, J, O, U, OO, PP are large compared to the others ($\sim$2 \arcsec~in diameter or $\sim$ 17 kpc). A few appear as disks (or inclined disks) and some others are bulge-dominated, but most have irregular morphologies. In Figure \ref{fig:morph} we compare the Gini, M$_{20}$, and concentration parameters of these galaxies (labeled orange stars) compared to high redshift star-forming galaxies, emission-line selected galaxies at $z=1.5$ (blue crosses) and LBGs at $z=4.0$ (magenta diamonds) from \cite{Lotz06}, and low redshift LBAs at $z=0.2$ (black open circles) from \cite{rod}. In the left panel, the left hatched region describes mergers and the upper right hatched region describes bulge-dominated galaxies; in both panels, the red and navy shaded regions mark the simulated parameters for galaxies following de Vaucouleurs and exponential light profiles, respectively. We note that the \cite{Lotz06} and \cite{rod} galaxy morphologies are determined from the rest frame FUV and we use rest-frame NUV (observed V$_{606}$-band), since B$_{435}$-band images were not available for many of the galaxies. We note that the morphology parameters do not change significantly when using the B$_{435}$-band data for the subset of galaxies with available B$_{435}$-band images. Our LBGs occupy the same parameter space (with similar scatter and range) as the $z=1.5$ emission-line selected star-forming galaxies and $z\sim0.2$ LBAs. 19 of the 50 LBGs resemble ``mergers'', according to the G-M$_{20}$ plane. 

 \begin{figure}
\begin{center}
  \includegraphics[width=3.5in]{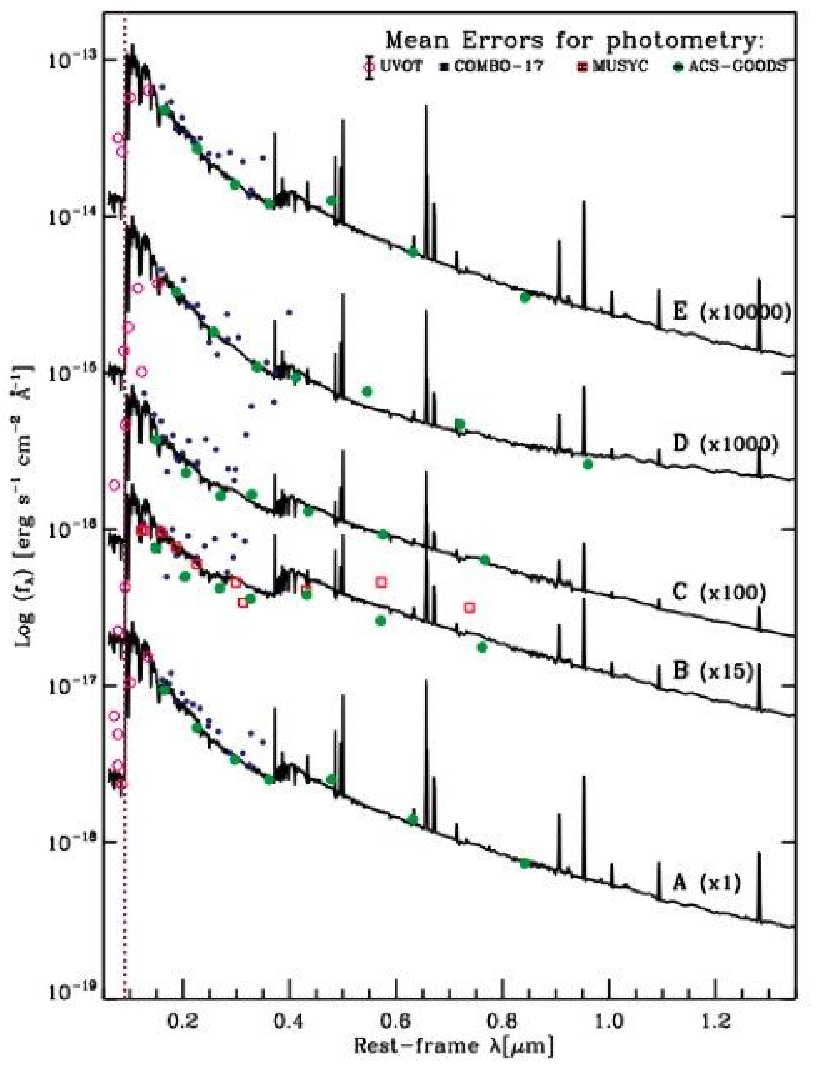}
  \end{center}
  \vspace{-0.2in}
  \caption{We compare the best-fit SED for each galaxy (LBGs A thru E are shown here; SEDs for the other 45 LBGs are available in the online version) with photometric data (COMBO-17 appears as small blue points, MUSYC as magenta squares, ACS-GOODS as larger green points and UVOT in open magenta circles). Observed data have been shifted by ($1+z$) factor into rest frame. Purple dotted line marks the rest-frame Lyman limit at 912\AA. SEDs are labeled on right with identifying names. Y-axis shows Log f$_\lambda$ in [erg s$^{-1}$ cm$^{-2}$ \AA$^{-1}$], displaced by some factor (quoted to the right of the ID label) to display five SEDs together, without overlap. Photometric errors are usually smaller than the symbols (except for UVOT). We show typical mean uncertainties for each dataset at the top. }
   \label{fig:seds}
\vspace{0.05in}
   \end{figure}
\subsection{SEDs and Photometry}\label{sec:sed}
SEDs were fit using MUSYC, COMBO-17 and ACS-GOODS photometry -- we tried fitting all three sets of photometry together, combinations of any two sets and each set separately to determine the best-fit SED. We were not able to fit the SED for five of the galaxies: LBGs K,  O,  S,  BB,  or DD.  LBG K had both ACS-GOODS and COMBO-17 photometry and LBG O had both MUSYC and COMBO-17; in both cases, neither photometric dataset (together nor separately) could be fit to produce a single SED. LBGs S, BB, and DD had only COMBO-17 data, which also could not be fit by any SED. In Figure \ref{fig:seds}, we show the best-fit rest-frame SEDs for our LBG sample (LBGs A$-$E are shown in the paper, the rest are available electronically) with all available photometric data marked as colored points: ACS-GOODS in green, COMBO-17 in smaller blue, MUSYC as open red squares, and UVOT in open magenta circles. We note that the SEDs for LBGs OO and PP are more consistent with elliptical galaxies; as Figure \ref{fig:sel} indicates, the $z=1$ elliptical track is on the edge of our selection region.  

As discussed in \S\ref{sec:opt}, while offering additional information, we face challenges reconciling and interpreting photometry from different catalogs. For example, differences between photometry from different datasets depend on SExtractor segment maps. 
 In most cases, the photometry from COMBO-17, ACS-GOODS and MUSYC are fit well by the best-fit spectrum.  However, we note a few exceptional cases. LBG B is one of the more extended sources and it isn't clear that the different catalogs use consistent apertures for calculating photometry -- photometry from COMBO-17 gives brightest fluxes, MUSYC data and ACS-GOODS photometry are well fit by the simultaneous best-fit to all of these datasets, but deviate from each other at rest-frame wavelengths, $\lambda_{\rm rf} >4500$ \AA. LBG G appears brighter in the COMBO-17 catalog than in the ACS-GOODS catalog. The COMBO-17 data for LBGs F, W, II, KK and TT show some structure, that is not apparent in the broadband ACS-GOODS data and does not match the best-fit SED. The best-fit SED and ACS-GOODS data match COMBO-17 data in LBG VV for $\lambda_{\rm rf} <2000$ \AA, but then the COMBO-17 data appears to deviate. In LBGs WW and G the COMBO-17 data is somewhat brighter than the ACS-GOODS data and best-fit SED. While, LBG P was successfully fit with an SED, the significant scatter in the photometry (along with the GEMS optical image in Figure \ref{fig:mapsbw}) suggest that this candidate may be two separate galaxies. 

We note that fits including the NIR (including either MUSYC or ACS-GOODS) give more realistic results. For example, in LBGs Y, CC and XX, the COMBO-17 points match the best-fit SED well, but the lack of data for $\lambda_{\rm rf} > 4000$ \AA~leaves this regime unconstrained. This results in unrealistically large derived stellar masses, Log(M$_*$)$=10.7$, 11.1 and 11.0.

\begin{figure}
\begin{center}
    \includegraphics[width=3.5in]{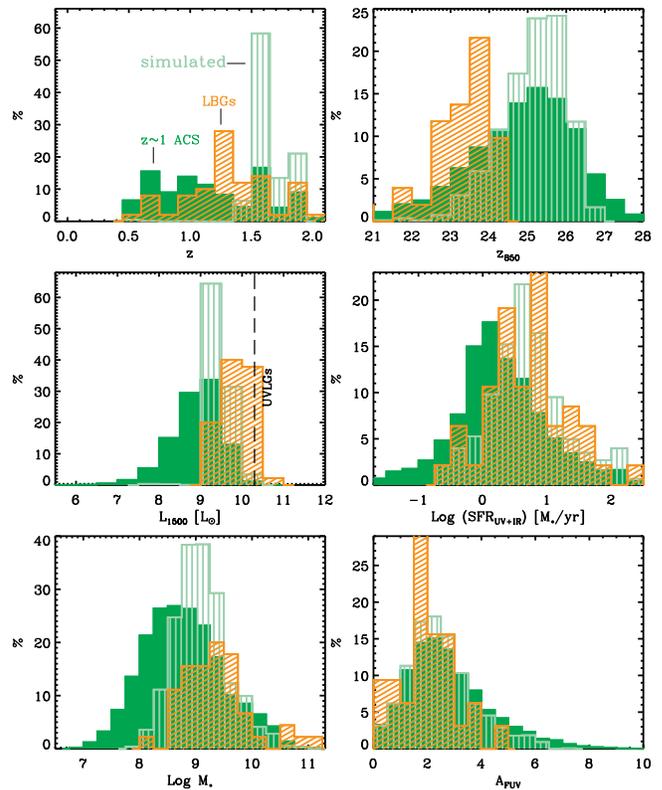}

    \end{center}
  \caption{We compare the distributions of observed and derived parameters for $z\sim1$ ACS (green solid region), simulated LBGs (light green vertically striped region), and UVOT-selected LBGs (orange shaded region): photometric redshift, apparent z$_{850}$ magnitude, rest-frame UV luminosity, total SFR, stellar mass and FUV dust attenuation (see text for details). In the UV luminosity distribution panel (middle row, first column), the selection criterion for $z\sim 0.2$ UVLGs has been marked (L$_{\rm FUV}=10^{10.3} {\rm L}_\odot$; see discussion in Section \ref{sec:discother}). Distributions have been normalized using the number of galaxies with valid measurements (\ie galaxies with fitting errors or missing data were not included in the normalization). While brighter than the simulated or ACS-GOODS sample, the LBGs share similar properties (L$_{1500}$, SFR, and mass) as the simulated sample.}
   \label{fig:dist}
\vspace{0.05in}
   \end{figure}
   
\begin{figure*}
\begin{center}
  \includegraphics[width=6.5in]{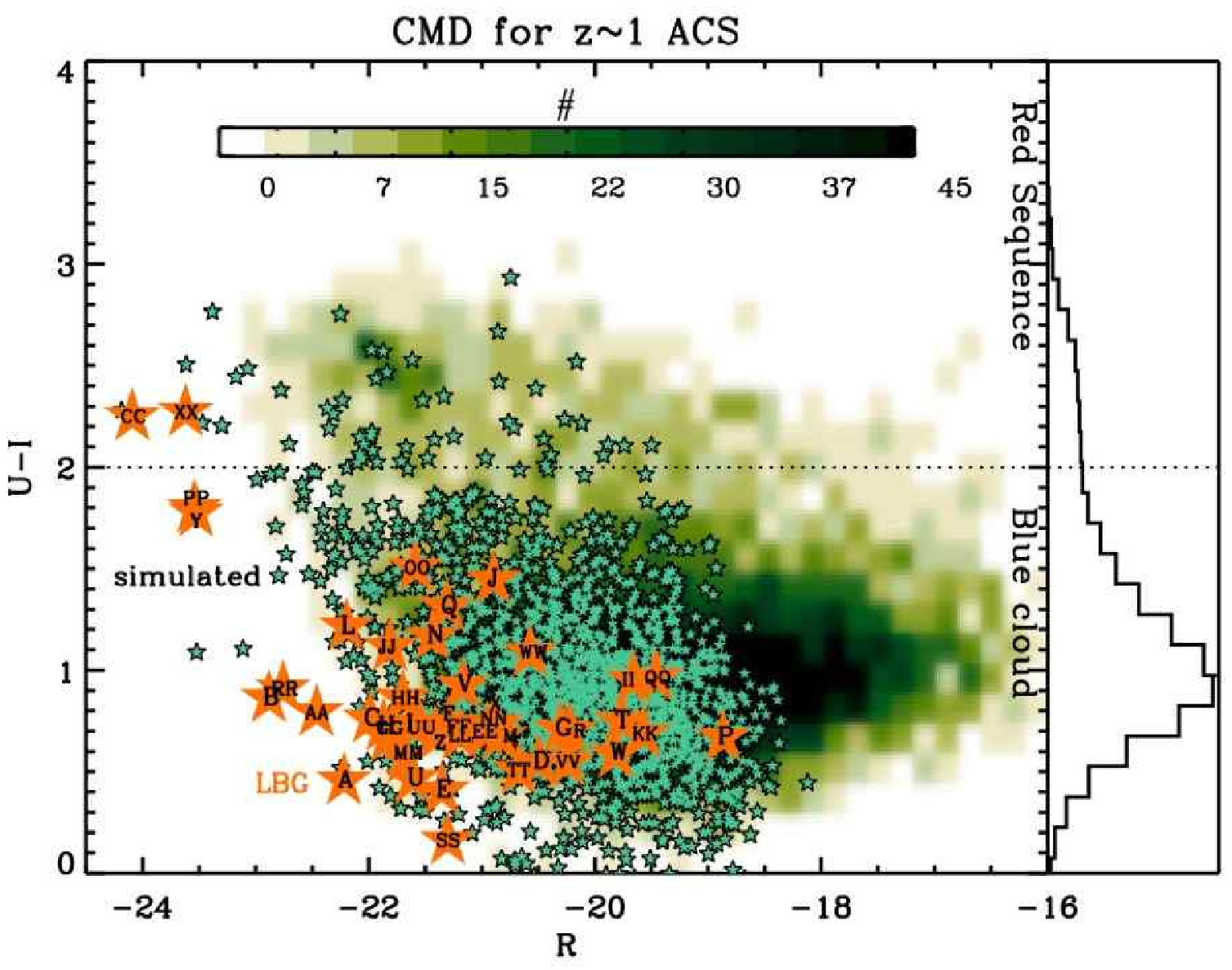}
    \end{center}
  \caption{The UVOT-selected LBGs (orange stars, labeled by their name) reside in the blue cloud (U$-$I $<$ 2, marked by the dotted line) along with most of the simulated galaxies, shown as light green stars (only $\sim3\%$ are in the ``red sequence''), as shown by this rest-frame color-magnitude diagram. The distribution of the entire $z\sim1$ ACS \citep{Dahlen2010} sample, is shown as the dark green background. Histogram on right side shows the U$-$I color distribution for the $z\sim1$ ACS galaxies.}
   \label{fig:cmd}
\vspace{0.05in}
   \end{figure*}

 \subsection{UVOT-selected LBGs compared to other samples}\label{sec:discother}
The distributions of derived (\eg redshift,  rest-frame L$_{1500}$, UV$+$IR star formation rate, stellar mass and dust attenuation in the FUV) and observed (\eg z$_{850}$ magnitude) properties for the LBGs are shown in Figure \ref{fig:dist} by orange shaded regions and the properties for the two comparison samples, $z\sim1$ ACS sample and simulated LBGS, are shown in solid dark green and vertically striped light green regions. The LBGs appear to be have slightly higher SFRs [$\langle \rm Log~SFR/({\rm M_\odot~yr^{-1}})\rangle\sim0.7 (\pm 0.6)$] and stellar masses [$\langle \rm Log~M_*/M_\odot\rangle \sim 9.4 (\pm 0.6)$] compared to the $z\sim1$ ACS sample [$\langle \rm Log~SFR/({\rm M_\odot~yr^{-1}})\rangle\sim0.2 (\pm 0.7)$ and $\langle \rm Log~M_*/M_\odot\rangle \sim 8.7 (\pm 0.7)$]. Although the distributions are quite broad for both samples, a K-S test indicates that the distributions of the LBG sample differ from those of the $z\sim1$ ACS sample in both SFR (K-S probability $\sim3\times 10^{-5}$) and M$_*$ (K-S probability $\sim 4\times10^{-7}$). However, compared to the simulated LBGS, the LBGs do have more similar distributions in stellar mass (K-S probability $\sim 0.008$) and extremely similar distributions in SFR$_{\rm UV+IR}$ (K-S probability $\sim 0.9$) while being $\sim$2.5 magnitudes brighter in z$_{850}$(due to selection effects). The redshift distribution of the simulated LBGs looks different from either the $z\sim1$ ACS sample or the observed LBGs -- with a tighter distribution peaked at $z\sim1.8$.

We compare our LBG sample with other studies. LBAs \citep[see][]{Choopes, Heckman05} are UV luminous galaxies (UVLGs; L$_{1500}>10^{10.3}~{\rm L_\odot}$) with high galaxy-wide mean FUV surface brightnesses (I$_{\rm FUV}> 10^{9}~\rm L_\odot~{\rm kpc}^{-2}$) at $z\sim0.2$. Based on these requirements, LBGs A and RR are both UVLGs and LBAs. The others are less UV luminous and have lower FUV surface brightnesses, with $\langle \rm L_{1500}\rangle=10^{9.9}~L_\odot$ and $\langle \rm I_{\rm FUV} \rangle=10^{8.7}~L_\odot~{\rm kpc}^{-2}$. LBGs selected in this paper have an order of magnitude lower SFRs (as derived by UV$+$IR; $-0.7<{\rm Log ~SFR/(M_\odot~yr^{-1}})<2.3$) compared to either $z\sim 3$ LBGs ($0.5<{\rm Log~SFR/(M_\odot ~yr^{-1}})<2.5$) or $z\approx0.2$ LBAs ($0.5<{\rm Log~SFR/(M_\odot~yr^{-1}})<2.0$). The stellar masses are in similar range, although the LBGs in this study have masses extending to a slightly wider range ($8.1<{\rm Log~M_*/M}_\odot< 11.1$) than $z\sim3$ LBGs ($9.5<{\rm Log ~M_* /M}_\odot<11.0$) or $z\sim0.2$ LBAs ($9.0<{\rm Log~M_*/M}_\odot<10.7$). Meanwhile the FUV attenuations in these LBGs are similar but extend to slightly larger ${\rm A_{FUV}}$ values ($0<{\rm A}_{\rm FUV}<5$) than $z\sim3$ LBGs (0$<{\rm A_{FUV}}<3$) or $z\sim0.2$ LBAs ($0<\rm{A_{FUV}}<2$). 

\cite{Burgarella2006} studied the IR properties of \GALEX-selected $z\sim1$ LBGs and find that $\sim95$\% of the IR-detected LBGs could be classified as Luminous Infrared Galaxies (LIRGs; \ie have L$_{\rm IR}> 10^{11} {\rm L}_\odot$). We also find that a significant fraction of our LBGs are LIRGs, though not as high as that found by \cite{Burgarella2006}: out of the 34 LBGs with MIPS counterparts, 11 have L$_{\rm IR}>10^{11}~L_\odot$ (\ie $\sim32\%$ of the IR-detected LBGs are LIRGs). 

We compare the color magnitude diagram of our LBGs to the comparison samples ($z\sim1$ ACS sample is shown as a dark green density distribution and simulated LBGs are marked as light green stars) in Figure \ref{fig:cmd}. In general, the LBGs (both observed and simulated) appear bluer and more luminous, as expected from the UVOT selection (based on color, see Figure \ref{fig:sel}, and $U>$24.5). 

As discussed in Section \ref{sec:comp}, $\sim$7\% of the simulated LBGs do appear redder and with SEDs dominated by older stellar populations, and from this figure we find that 3\% are on the ``red sequence'' (with U$-$I $>2.0$). We have 2 galaxies (LBGs CC and XX) that appear in the red sequence and 3 others (LBGs Y, OO, and  PP) that appear close to the red sequence. LBGs Y, CC and XX don't have available NIR data, leaving $\lambda_{\rm rf} \geq 4000$ \AA~unconstrained -- these galaxies all have similar SEDs which show a steep upturn for $\lambda_{\rm rf} > 4000$ \AA. For that reason, they also have unusually large derived stellar masses (see discussion in \S\ref{sec:sed}) and the R magnitudes are likely overestimated from the unconstrained SEDs. LBGs OO and PP are unusual from the other LBGs; they appears both redder and brighter than the other LBGs, and their SEDs, as noted in \S\ref{sec:sed}, do show large 4000 \AA~breaks and appear dominated by older stellar populations. LBG PP does appear unusually bright in R, but we note that the redshift for this galaxy was spectroscopically determined. While there are challenges with determining spectroscopic redshifts in this redshift range and it is possible that the redshift for this object is an overestimate; however, photometric redshifts from both COMBO-17 and ACS-GOODS are consistent with the spectroscopic redshift.

\section{Discussion and Summary}\label{sec:end}
In this paper, we used \Swift~UVOT to identify and select LBG candidates. Based on redshifts determined from other multiwavelength catalogs (\eg MUSYC, COMBO-17 and ACS-GOODs), we determined that 50 candidates were LBGs. We use these multiwavelength catalogs to fit SEDs to determine stellar masses, k-corrected absolute magnitudes and star formation rates. Using available MIPS data for 40 of the galaxies, we also determined UV$+$IR SFRs and FUV dust attenuations, A$_{\rm FUV}$. GEMS and ACS-GOODS images offer high resolution optical morphologies of these galaxies. From our study we have determined that \Swift~UVOT can select 0.5$<z<$ 2.0 LBGs within deep observed fields, but redshift confirmation or further photometric data is required as we found contamination (in $\sim10-15$\% of the candidates) from low redshift interlopers.
 
We found that the UVOT-selected LBGs have similar morphologies to $z=0.2$ LBAs from \cite{rod}, $z=1.5$ emission line selected star-forming galaxies, and $z=4$ LBGs from \cite{Lotz06} (Figure \ref{fig:morph}). These LBGs also have similar values for SFR, stellar mass and FUV dust attenuation, but span larger ranges, compared with $z\sim3$ LBGs and $z\sim0.2$ LBAs. However, compared to the $z\sim1$ ACS sample, these LBGs are bluer and brighter and have slightly higher stellar masses, and marginally higher total SFRs. 

We used best-fit SEDs to simulate UVOT photometry for the $z\sim1$ ACS sample and selected a sample of simulated LBGs based on the same criteria applied on the observed UVOT-selected LBGS. Despite including fainter [$\langle {\rm z_{850} (simulated)} \rangle \approx \langle {\rm z_{850} (observed)} \rangle +2.5$ mag] galaxies, the simulated sample does exhibit similar properties as the observed UVOT-selected LBG sample: slightly more massive, more UV-luminous, and slightly higher SFRs compared to the $z\sim1$ ACS sample (Figures \ref{fig:dist} and \ref{fig:cmd}). The LBGs (observed and simulated) have similar dust attenuation compared to the $z\sim1$ ACS comparison sample and the amount of FUV attenuation is not as low as in the $z\sim3$ LBGs or $z\sim0.2$ LBAs. We find that  $\sim32\%$ of the IR-detected LBGs are LIRGs. This fraction is not as high as what \cite{Burgarella2006} found for their $z\sim1$ \GALEX-selected LBGs, but it is a significant fraction of the LBGs. Red (U$-$I$>2$) galaxies were found in both the simulated LBG sample ($\sim 3$\%) and the observed LBG sample (2 galaxies, or $\sim4$\%). 

This research investigates the viability of using \Swift~UVOT to select intermediate redshift ($z=0.5 - 2$) LBGs. Until recently, there was little opportunity to study LBGs in this redshift desert. However, with the WFC3 upgrades to \HST~the new ultraviolet filters have been used to efficiently select hundreds of LBGs in this regime \citep{Hathi2010, Oesch2010}. While WFC3 has higher sensitivity and can select fainter LBGs more efficiently, its FOV is small; GALEX data has larger sky coverage, yet it suffers source confusion from low spatial resolution and does not have multiple NUV filters. We have shown that UVOT can be valuable for studying the bright end of the LBG sample. Having compared with a simulated LBG sample, we learn that the fainter, simulated LBGs appear to resemble the observed, brighter LBGs in important properties (\eg stellar mass, SFR, dust attenuation). According to statistical studies of the contribution of UV-luminous galaxies to the SFR density \citep{schimi04}, the number of LBGs decreases rapidly in the low redshift Universe ($z<2$). Since UVOT has covered a larger fraction of the sky than WFC3, UVOT data has the potential to recover a larger (albeit, shallow) sample of LBGs in this redshift range. There are several other deep UVOT fields (40 GRB fields with UV exposure times of $\geq$ 200 ks) for which a similar analysis can be done to further study LBGs between $0.5<z<2$. 

As future missions, such as James Webb Space Telescope (JWST), push to discover more distant (rest-frame UV-selected) galaxies, it becomes more important to understand the selection of these galaxies in the relatively nearby Universe, as well as how their properties compare with the higher redshift ($z>2$) population.     
\begin{acknowledgments}
This research was supported by an appointment to the NASA Postdoctoral Program at the Goddard Space Flight Center, administered by Oak Ridge Associated Universities through a contract with NASA. We acknowledge support from NASA Astrophysics Data Analysis grant NNX09AC87G. This work is sponsored at  PSU by NASA contract NAS5-00136. We appreciate and acknowledge the members of the \Swift~ instrument and science team for their efforts, particularly Dean Hinshaw and Scott Koch for their work on the UVOT catalog and images. 
\end{acknowledgments}

\bibliographystyle{apj}
\bibliography{apj-jour,antara_refs,library_mod}

\begin{thebibliography}{80}
\expandafter\ifx\csname natexlab\endcsname\relax\def\natexlab#1{#1}\fi

\bibitem[{{Abraham} {et~al.}(1996){Abraham}, {van den Bergh}, {Glazebrook},
  {Ellis}, {Santiago}, {Surma}, \& {Griffiths}}]{Abraham96}
{Abraham}, R.~G., {et~al.} 1996, \apjs, 107, 1

\bibitem[{{Abraham} {et~al.}(2003){Abraham}, {van den Bergh}, \&
  {Nair}}]{Abraham2003}
{Abraham}, R.~G., {van den Bergh}, S., \& {Nair}, P. 2003, \apj, 588, 218

\bibitem[{{Adelberger} {et~al.}(2003){Adelberger}, {Steidel}, {Shapley}, \&
  {Pettini}}]{Adel03}
{Adelberger}, K.~L., {Steidel}, C.~C., {Shapley}, A.~E., \& {Pettini}, M. 2003,
  \apj, 584, 45

\bibitem[{{Adelberger} {et~al.}(2005)}]{adel05}
{Adelberger}, K.~L., {et~al.} 2005, \apjl, 620, L75

\bibitem[{Arnouts {et~al.}(2005)}]{Arnouts04}
Arnouts, S., {et~al.} 2005, \apjl, 619, L43

\bibitem[{{Basu-Zych} {et~al.}(2007)}]{me2}
{Basu-Zych}, A.~R., {et~al.} 2007, \apjs, 173, 457

\bibitem[{{Basu-Zych} {et~al.}(2009{\natexlab{a}})}]{me-ifu}
---. 2009{\natexlab{a}}, \apjl, 699, L118

\bibitem[{{Basu-Zych} {et~al.}(2009{\natexlab{b}})}]{me-env}
---. 2009{\natexlab{b}}, \apj, 699, 1307

\bibitem[{{Bertin} \& {Arnouts}(1996)}]{sex}
{Bertin}, E., \& {Arnouts}, S. 1996, \aaps, 117, 393

\bibitem[{{Blanton} \& {Roweis}(2007)}]{kcorrect2}
{Blanton}, M.~R., \& {Roweis}, S. 2007, \aj, 133, 734

\bibitem[{{Bouwens} {et~al.}(2006){Bouwens}, {Illingworth}, {Blakeslee}, \&
  {Franx}}]{Bouwens}
{Bouwens}, R.~J., {Illingworth}, G.~D., {Blakeslee}, J.~P., \& {Franx}, M.
  2006, \apj, 653, 53

\bibitem[{{Bouwens} {et~al.}(2008){Bouwens}, {Illingworth}, {Franx}, \&
  {Ford}}]{Bouwens08}
{Bouwens}, R.~J., {Illingworth}, G.~D., {Franx}, M., \& {Ford}, H. 2008, \apj,
  686, 230

\bibitem[{{Bouwens} {et~al.}(2009)}]{Bouwens09}
{Bouwens}, R.~J., {et~al.} 2009, \apj, 705, 936

\bibitem[{{Bouwens} {et~al.}(2010)}]{Bouwens2010}
---. 2010, in press (arXiv:1006.4360)

\bibitem[{{Breeveld} {et~al.}(2011){Breeveld}, {Landsman}, {Holland}, {Roming},
  {Kuin}, \& {Page}}]{Breeveld11}
{Breeveld}, A.~A., {et~al.} 2011, in press (arXiv:1102.4717)

\bibitem[{{Breeveld} {et~al.}(2010)}]{Breeveld10}
---. 2010, \mnras, 406, 1687

\bibitem[{{Brown} {et~al.}(2010)}]{Brown10}
{Brown}, P.~J., {et~al.} 2010, \apj, 721, 1608

\bibitem[{{Bruzual} \& {Charlot}(2003)}]{BC03}
{Bruzual}, G., \& {Charlot}, S. 2003, MNRAS, 344, 1000

\bibitem[{{Burgarella} {et~al.}(2005){Burgarella}, {Buat}, \&
  {Iglesias-P{\'a}ramo}}]{Burgarella2005a}
{Burgarella}, D., {Buat}, V., \& {Iglesias-P{\'a}ramo}, J. 2005, \mnras, 360,
  1413

\bibitem[{Burgarella {et~al.}(2006)Burgarella, Perez-Gonzalez, Tyler, Rieke,
  Buat, Takeuchi, Lauger, Arnouts, Ilbert, Barlow, Bianchi, Madore, Malina,
  Szalay, \& Yi}]{Burgarella2006}
Burgarella, D., {et~al.} 2006, \aap, 450, 69

\bibitem[{{Calzetti} {et~al.}(1994){Calzetti}, {Kinney}, \&
  {Storchi-Bergmann}}]{Calzetti94}
{Calzetti}, D., {Kinney}, A.~L., \& {Storchi-Bergmann}, T. 1994, \apj, 429, 582

\bibitem[{{Castelli} \& {Kurucz}(2003)}]{ckstars}
{Castelli}, F., \& {Kurucz}, R.~L. 2003, in IAU Symposium, Vol. 210, Modelling
  of Stellar Atmospheres, ed. {N.~Piskunov, W.~W.~Weiss, \& D.~F.~Gray}, 20P

\bibitem[{{Chary}(2007)}]{Charyproc}
{Chary}, R. 2007, in Astronomical Society of the Pacific Conference Series,
  Vol. 380, Deepest Astronomical Surveys, ed. {J.~Afonso, H.~C.~Ferguson,
  B.~Mobasher, \& R.~Norris}, 375

\bibitem[{{Chary} \& {Elbaz}(2001)}]{CharyElbaz}
{Chary}, R., \& {Elbaz}, D. 2001, \apj, 556, 562

\bibitem[{{Chary} {et~al.}(2004)}]{Chary2004}
{Chary}, R., {et~al.} 2004, \apjs, 154, 80

\bibitem[{{Conselice}(2003)}]{Conselice03}
{Conselice}, C.~J. 2003, ApJS, 147, 1

\bibitem[{{Conselice} {et~al.}(2000){Conselice}, {Bershady}, \&
  {Jangren}}]{Conselice2000}
{Conselice}, C.~J., {Bershady}, M.~A., \& {Jangren}, A. 2000, \apj, 529, 886

\bibitem[{Dahlen {et~al.}(2007)Dahlen, Mobasher, Dickinson, Ferguson,
  Giavalisco, Kretchmer, \& Ravindranath}]{DahlenTomas2007}
Dahlen, T., {et~al.} 2007, \apj, 654, 172

\bibitem[{{Dahlen} {et~al.}(2010){Dahlen}, {Mobasher}, {Dickinson}, {Ferguson},
  {Giavalisco}, {Grogin}, {Guo}, {Koekemoer}, {Lee}, {Lee}, {Nonino}, {Riess},
  \& {Salimbeni}}]{Dahlen2010}
{Dahlen}, T., {et~al.} 2010, \apj, 724, 425

\bibitem[{{Elbaz} {et~al.}(2010)}]{Elbaz2010}
{Elbaz}, D., {et~al.} 2010, \aap, 518, L29

\bibitem[{{Fioc} \& {Rocca-Volmerange}(1997)}]{peg}
{Fioc}, M., \& {Rocca-Volmerange}, B. 1997, \aap, 326, 950

\bibitem[{{Gehrels} {et~al.}(2004)}]{Gehrels04}
{Gehrels}, N., {et~al.} 2004, \apj, 611, 1005

\bibitem[{{Giavalisco}(2002)}]{giarev}
{Giavalisco}, M. 2002, \araa, 40, 579

\bibitem[{{Giavalisco} \& {Dickinson}(2001)}]{gia}
{Giavalisco}, M., \& {Dickinson}, M. 2001, \apj, 550, 177

\bibitem[{{Giavalisco} {et~al.}(2004{\natexlab{a}})}]{giaGOODS}
{Giavalisco}, M., {et~al.} 2004{\natexlab{a}}, \apjl, 600, L93

\bibitem[{{Giavalisco} {et~al.}(2004{\natexlab{b}}){Giavalisco}, {Dickinson},
  {Ferguson}, {Ravindranath}, {Kretchmer}, {Moustakas}, {Madau}, {Fall},
  {Gardner}, {Livio}, {Papovich}, {Renzini}, {Spinrad}, {Stern}, \&
  {Riess}}]{gia2004}
---. 2004{\natexlab{b}}, \apjl, 600, L103

\bibitem[{{Gon{\c c}alves} {et~al.}(2010){Gon{\c c}alves}, {Basu-Zych},
  {Overzier}, {Martin}, {Law}, {Schiminovich}, {Wyder}, {Mallery}, {Rich}, \&
  {Heckman}}]{tsg}
{Gon{\c c}alves}, T.~S., {et~al.} 2010, \apj, 724, 1373

\bibitem[{{Hathi} {et~al.}(2010){Hathi}, {Ryan}, {Cohen}, {Yan}, {Windhorst},
  {McCarthy}, {O'Connell}, {Koekemoer}, {Rutkowski}, {Balick}, {Bond},
  {Calzetti}, {Disney}, {Dopita}, {Frogel}, {Hall}, {Holtzman}, {Kimble},
  {Paresce}, {Saha}, {Silk}, {Trauger}, {Walker}, {Whitmore}, \&
  {Young}}]{Hathi2010}
{Hathi}, N.~P., {et~al.} 2010, \apj, 720, 1708

\bibitem[{{Heckman} {et~al.}(2005)}]{Heckman05}
{Heckman}, T.~M., {et~al.} 2005, \apjl, 619, L35

\bibitem[{{Hoopes} {et~al.}(2007)}]{Choopes}
{Hoopes}, C.~G., {et~al.} 2007, \apjs, 173, 441

\bibitem[{{Hopkins} \& {Beacom}(2006)}]{HB06}
{Hopkins}, A.~M., \& {Beacom}, J.~F. 2006, \apj, 651, 142

\bibitem[{{Hoversten} {et~al.}(2009)}]{Hoversten10}
{Hoversten}, E.~A., {et~al.} 2009, \apj, 705, 1462

\bibitem[{{Hoversten} {et~al.}(2011){Hoversten}, {Gronwall}, {Vanden Berk},
  {Basu-Zych}, {Breeveld}, {Brown}, {Kuin}, {Page}, {Roming}, \&
  {Siegel}}]{Hoversten11}
---. 2011, \aj, 141, 205

\bibitem[{{Kennicutt}(1998)}]{Kennicutt}
{Kennicutt}, R.~C. 1998, \araa, 36, 189

\bibitem[{{Kewley} {et~al.}(2001){Kewley}, {Dopita}, {Sutherland}, {Heisler},
  \& {Trevena}}]{Kewley01}
{Kewley}, L.~J., {Dopita}, M.~A., {Sutherland}, R.~S., {Heisler}, C.~A., \&
  {Trevena}, J. 2001, \apj, 556, 121

\bibitem[{{Kroupa}(2001)}]{Kroupa}
{Kroupa}, P. 2001, \mnras, 322, 231

\bibitem[{{Lehmer} {et~al.}(2008)}]{Lehmer08}
{Lehmer}, B.~D., {et~al.} 2008, \apj, 681, 1163

\bibitem[{{Lotz} {et~al.}(2006){Lotz}, {Madau}, {Giavalisco}, {Primack}, \&
  {Ferguson}}]{Lotz06}
{Lotz}, J.~M., {Madau}, P., {Giavalisco}, M., {Primack}, J., \& {Ferguson},
  H.~C. 2006, \apj, 636, 592

\bibitem[{{Lotz} {et~al.}(2004){Lotz}, {Primack}, \& {Madau}}]{Lotz04}
{Lotz}, J.~M., {Primack}, J., \& {Madau}, P. 2004, \aj, 128, 163

\bibitem[{{Luo} {et~al.}(2008)}]{Luo}
{Luo}, B., {et~al.} 2008, \apjs, 179, 19

\bibitem[{Ly {et~al.}(2009)Ly, Malkan, Treu, Woo, Currie, Hayashi, Kashikawa,
  Motohara, Shimasaku, \& Yoshida}]{Ly2009}
Ly, C., {et~al.} 2009, \apj, 697, 1410

\bibitem[{{Madau}(1995)}]{Madau95}
{Madau}, P. 1995, \apj, 441, 18

\bibitem[{{Madau} {et~al.}(1996){Madau}, {Ferguson}, {Dickinson}, {Giavalisco},
  {Steidel}, \& {Fruchter}}]{Madau96}
{Madau}, P., {et~al.} 1996, \mnras, 283, 1388

\bibitem[{{Martin} {et~al.}(2005){Martin}, {Fanson}, {Schiminovich},
  {Morrissey}, {Friedman}, {Barlow}, {Conrow}, {Grange}, {Jelinsky},
  {Milliard}, {Siegmund}, {Bianchi}, {Byun}, {Donas}, {Forster}, {Heckman},
  {Lee}, {Madore}, {Malina}, {Neff}, {Rich}, {Small}, {Surber}, {Szalay},
  {Welsh}, \& {Wyder}}]{galex}
{Martin}, D.~C., {et~al.} 2005, \apjl, 619, L1

\bibitem[{{Mori} \& {Umemura}(2007)}]{Mori}
{Mori}, M., \& {Umemura}, M. 2007, in Engineering and Science, Vol.~24, EAS
  Publications Series, 221

\bibitem[{{Oesch} {et~al.}(2010){Oesch}, {Bouwens}, {Carollo}, {Illingworth},
  {Magee}, {Trenti}, {Stiavelli}, {Franx}, {Labb{\'e}}, \& {van
  Dokkum}}]{Oesch2010}
{Oesch}, P.~A., {et~al.} 2010, \apjl, 725, L150

\bibitem[{{Ouchi} {et~al.}(2005)}]{ouchi}
{Ouchi}, M., {et~al.} 2005, \apjl, 620, L1

\bibitem[{{Overzier} {et~al.}(2010){Overzier}, {Heckman}, {Schiminovich},
  {Basu-Zych}, {Gon{\c c}alves}, {Martin}, \& {Rich}}]{Overzier2010}
{Overzier}, R.~A., {et~al.} 2010, \apj, 710, 979

\bibitem[{{Overzier} {et~al.}(2008)}]{rod}
---. 2008, \apj, 677, 37

\bibitem[{{Papovich} {et~al.}(2005){Papovich}, {Dickinson}, {Giavalisco},
  {Conselice}, \& {Ferguson}}]{Papovich2005}
{Papovich}, C., {Dickinson}, M., {Giavalisco}, M., {Conselice}, C.~J., \&
  {Ferguson}, H.~C. 2005, \apj, 631, 101

\bibitem[{{Partridge} \& {Peebles}(1967)}]{PP1}
{Partridge}, R.~B., \& {Peebles}, P.~J.~E. 1967, \apj, 147, 868

\bibitem[{{Pei}(1992)}]{Pei92}
{Pei}, Y.~C. 1992, \apj, 395, 130

\bibitem[{{Pettini} {et~al.}(2002)}]{pettini02}
{Pettini}, M., {et~al.} 2002, \apj, 569, 742

\bibitem[{{Poole} {et~al.}(2008)}]{Poole08}
{Poole}, T.~S., {et~al.} 2008, \mnras, 383, 627

\bibitem[{{Renzini} \& {Daddi}(2009)}]{Renzini2009}
{Renzini}, A., \& {Daddi}, E. 2009, The Messenger, 137, 41

\bibitem[{{Rieke} {et~al.}(2004){Rieke}, {Young}, {Engelbracht}, {Kelly},
  {Low}, {Haller}, {Beeman}, {Gordon}, {Stansberry}, {Misselt}, {Cadien},
  {Morrison}, {Rivlis}, {Latter}, {Noriega-Crespo}, {Padgett}, {Stapelfeldt},
  {Hines}, {Egami}, {Muzerolle}, {Alonso-Herrero}, {Blaylock}, {Dole}, {Hinz},
  {Le Floc'h}, {Papovich}, {P{\'e}rez-Gonz{\'a}lez}, {Smith}, {Su}, {Bennett},
  {Frayer}, {Henderson}, {Lu}, {Masci}, {Pesenson}, {Rebull}, {Rho}, {Keene},
  {Stolovy}, {Wachter}, {Wheaton}, {Werner}, \& {Richards}}]{MIPS}
{Rieke}, G.~H., {et~al.} 2004, \apjs, 154, 25

\bibitem[{{Rix} {et~al.}(2004){Rix}, {Barden}, {Beckwith}, {Bell}, {Borch},
  {Caldwell}, {H{\"a}ussler}, {Jahnke}, {Jogee}, {McIntosh}, {Meisenheimer},
  {Peng}, {Sanchez}, {Somerville}, {Wisotzki}, \& {Wolf}}]{gems}
{Rix}, H., {et~al.} 2004, \apjs, 152, 163

\bibitem[{{Roming} {et~al.}(2005)}]{Roming05}
{Roming}, P.~W.~A., {et~al.} 2005, Space Science Reviews, 120, 95

\bibitem[{Schiminovich {et~al.}(2005)}]{schimi04}
Schiminovich, D., {et~al.} 2005, \apjl, 619, L47

\bibitem[{{Silverman} {et~al.}(2008){Silverman}, {Mainieri}, {Lehmer},
  {Alexander}, {Bauer}, {Bergeron}, {Brandt}, {Gilli}, {Hasinger}, {Schneider},
  {Tozzi}, {Vignali}, {Koekemoer}, {Miyaji}, {Popesso}, {Rosati}, \&
  {Szokoly}}]{Silverman}
{Silverman}, J.~D., {et~al.} 2008, \apj, 675, 1025

\bibitem[{{Steidel} \& {Hamilton}(1992)}]{Steidel92}
{Steidel}, C.~C., \& {Hamilton}, D. 1992, \aj, 104, 941

\bibitem[{{Steidel} \& {Hamilton}(1993)}]{Steidel93}
---. 1993, \aj, 105, 2017

\bibitem[{{Steidel} {et~al.}(1995){Steidel}, {Pettini}, \&
  {Hamilton}}]{Steidel95}
{Steidel}, C.~C., {Pettini}, M., \& {Hamilton}, D. 1995, \aj, 110, 2519

\bibitem[{{Steidel} {et~al.}(2000)}]{Steidel00}
{Steidel}, C.~C., {et~al.} 2000, \apj, 532, 170

\bibitem[{{Taylor} {et~al.}(2009)}]{musycK}
{Taylor}, E.~N., {et~al.} 2009, \apjs, 183, 295

\bibitem[{Vanzella {et~al.}(2009)Vanzella, Renzini, Giavalisco, Fosbury,
  Dickinson, Cristiani, Nonino, Kuntschner, Popesso, Rosati, Stern, Cesarsky,
  \& Ferguson}]{Vanzella2009}
Vanzella, E., {et~al.} 2009, ApJ, 695, 1163

\bibitem[{{Wolf} {et~al.}(2008){Wolf}, {Hildebrandt}, {Taylor}, \&
  {Meisenheimer}}]{Wolf2008}
{Wolf}, C., {Hildebrandt}, H., {Taylor}, E.~N., \& {Meisenheimer}, K. 2008,
  \aap, 492, 933

\bibitem[{{Wolf} {et~al.}(2004)}]{combo17}
{Wolf}, C., {et~al.} 2004, \aap, 421, 913

\bibitem[{{Wuyts} {et~al.}(2008)}]{FIREWORKS}
{Wuyts}, S., {et~al.} 2008, \apj, 682, 985

\bibitem[{{Zamojski} {et~al.}(2007)}]{Zamojski}
{Zamojski}, M.~A., {et~al.} 2007, \apjs, 172, 468

\end{thebibliography}

\end{document}